\pdfoutput=1
\documentclass[journal]{IEEEtran}

\usepackage[T1]{fontenc}

\ifCLASSINFOpdf
\else
\fi

\usepackage{graphicx}
\usepackage{textcomp}
\usepackage{float}

\def\BibTeX{{\rm B\kern-.05em{\sc i\kern-.025em b}\kern-.08em
    T\kern-.1667em\lower.7ex\hbox{E}\kern-.125emX}}

\usepackage[numbers,sort&compress]{natbib}
\usepackage[utf8]{inputenc} 
\usepackage[T1]{fontenc}    
\usepackage{hyperref}       
\usepackage{url}            
\usepackage{booktabs}       
\usepackage{amsfonts}       
\usepackage{nicefrac}       
\usepackage{microtype}      
\usepackage{cuted}
\usepackage{subcaption}
\usepackage{ifthen}
\usepackage{bbm}
\usepackage{amssymb}
\usepackage{algorithmicx}
\usepackage[cmex10]{amsmath}
\usepackage{amsthm}
\usepackage{mathtools}
\usepackage{bm}
\usepackage[inline, shortlabels]{enumitem}
\usepackage[electronic]{ifsym}
\usepackage{physics}

\usepackage{makecell,pbox,ragged2e,hhline}
\usepackage[table, x11names, svgnames]{xcolor}

\usepackage{color,diagbox,tabularx,multirow}

\usepackage{tikz}
\usetikzlibrary{fit, shapes.geometric, shapes.misc, arrows.meta, positioning, decorations.markings, matrix}
\tikzset{>={Latex[width=0.5mm,length=1.2mm]}}
\usepackage{smartdiagram}
\usesmartdiagramlibrary{additions}
\usepackage{pgfplots}
\pgfplotsset{compat=newest,    /pgfplots/ybar legend/.style={
    /pgfplots/legend image code/.code={
        \draw [#1] (0cm,-0.1cm) rectangle (0.4cm,0.1cm);},
   },}
\usepackage{varwidth}
\tikzset{every picture/.style={/utils/exec={\fontfamily{lmss}}}}
\tikzset{circ/.style = {circle, draw=black!100, fill=blue!25, thin, minimum height=3mm, minimum width=3mm, 
  inner sep=1.2mm},
  rect2/.style = {rectangle, draw=black!100, fill=cyan!18, thin, minimum height=7.5mm, minimum width=11mm, 
  inner sep=1.5mm},
  circ2/.style = {circle, draw=black!100, fill=red!35, thin, minimum size=7mm},
  outer/.style = {rounded corners=0.07cm, draw=DarkBlue, inner sep = 0mm, minimum height=6mm, ultra thick},
  outer2/.style = {rounded corners=0.2cm, draw=black!100, dashed, inner sep = 3mm},
  encode/.style = {trapezium, draw=black!100, fill=green!20, trapezium angle=75, shape border rotate=90, shift = {(-0.75, 0)}, minimum width=5mm, minimum height=11mm, align=center},
  decode/.style = {trapezium, draw=black!100, fill=green!20, trapezium angle=-75, shape border rotate=90, minimum width=5mm, minimum height=11mm, align=center},
  rect3/.style = {rectangle, draw=black!100, fill=cyan!1, thin, minimum height=7.5mm, minimum width=11mm, 
  inner sep=1.5mm},
  bluelines/.style={smooth, very thick, blue!40!gray},
  pre/.style={draw, rectangle, minimum height=6mm, minimum width=15mm, align=center,fill=gray!20, inner sep = 0mm},
  add/.style={draw, rectangle, minimum height=6mm, minimum width=20mm, align=center,fill=DarkBlue!20, inner sep = 0mm},
  msg/.style={draw, rectangle, minimum height=6mm, minimum width=32mm, align=center,fill=gray!20, inner sep = 0mm},
  par/.style={draw, rectangle, minimum height=6mm, minimum width=20mm, align=center,fill=gray!20, inner sep = 0mm},
  long/.style={short, text width=1.5cm},
  brace/.style={decorate, decoration={brace,amplitude=4pt,mirror,raise=1pt}},
  }

\usepackage{steinmetz}
\usepackage{xpatch}
\xpatchcmd{\phase}{#2}{\hspace{0.8pt}\vphantom{\scalebox{0.8}{\tiny{,}}}#2\hspace{1.4pt}}{}{}

\makeatletter
\newdimen\@widthOfTo%
\newdimen\@widthOfImplies%
\pgfmathsetmacro{\@scaleFactorImplies}{\@widthOfTo/\@widthOfImplies}%
\newcommand*{\ScaledImplies}{\mathrel{\raisebox{0.3ex}{\scalebox{\@scaleFactorImplies}{\ensuremath{\Longrightarrow}}}}}%
\makeatother

\newcolumntype{x}[1]{>{\centering\arraybackslash}p{#1}}
\newcolumntype{P}[1]{>{\centering\arraybackslash}p{#1}}
\newcolumntype{M}[1]{>{\centering\arraybackslash}m{#1}}

\usepackage{colortbl}
\definecolor{LightCyan}{rgb}{0.7,1,1}
\definecolor{LightGrey}{rgb}{0.95,0.95,0.95}
\definecolor{DarkBlue}{rgb}{0,0,0.4}
\definecolor{Yellow}{rgb}{1,1,0.6}

\DeclareMathOperator{\modrelu}{ModReLU}
\DeclareMathOperator{\crelu}{CReLU}
\DeclareMathOperator{\re}{Re}
\DeclareMathOperator{\im}{Im}

\let\originalleft\left
\let\originalright\right
\renewcommand{\left}{\mathopen{}\mathclose\bgroup\originalleft}
\renewcommand{\right}{\aftergroup\egroup\originalright}

\newcommand{\fc}{{f_{\mathrm{c}}}}
\renewcommand{\sc}{{s_{\mathrm{c}}}}
\renewcommand{\ss}{{s_{\mathrm{s}}}}
\newcommand{\srf}{{s_{\mathrm{RF}}}}

\newcommand{\bx}{{\bm x}}

\newcommand{\bxhat}{\bm{\widehat{x}}}

\newcommand{\bxideal}{{\bm x}_{\mathrm{ideal}} }
\newcommand{\thetaaug}{\theta_{\mathrm{aug}} }
\newcommand{\pdb}{P_{\mathrm{1\mkern 1mudB}}}

\theoremstyle{plain}

\theoremstyle{definition}

\theoremstyle{remark}

\usepackage{listings}
\lstdefinestyle{python}{
  belowcaptionskip=1\baselineskip,
  breaklines=true,
  frame=shadowbox,
  rulesepcolor=\color{gray},
  xleftmargin=\parindent,
  language=Python,
  showstringspaces=false,
  basicstyle=\footnotesize\ttfamily,
  keywordstyle=\bfseries\color{deepblue},
  moredelim=**[s][\color{blue}]{'''}{'''},
  commentstyle=\itshape\color{magenta},
  identifierstyle=\color{black},
  stringstyle=\color{red}
}
\lstdefinestyle{output}{
  belowcaptionskip=1\baselineskip,
  breaklines=true,
  frame=L,
  basicstyle=\footnotesize\ttfamily,
  xleftmargin=\parindent
}

\def\BibTeX{{\rm B\kern-.05em{\sc i\kern-.025em b}\kern-.08em T\kern-.1667em\lower.7ex\hbox{E}\kern-.125emX}}

\begin{document}

\title{Wireless Fingerprinting via Deep Learning: \\The Impact of Confounding Factors}

\author{Metehan~Cekic\IEEEauthorrefmark{1}\thanks{\IEEEauthorrefmark{1}Joint first authors.}\hspace{-1pt},~\IEEEmembership{Student Member,~IEEE,}
        Soorya~Gopalakrishnan\IEEEauthorrefmark{1}\hspace{-1pt},
        Upamanyu~Madhow,~\IEEEmembership{Fellow,~IEEE}%
\thanks{M.\ Cekic and U.\ Madhow are with the Department
of Electrical and Computer Engineering, University of California, Santa Barbara, CA 93106. (Email:\{metehancekic, madhow\}@ucsb.edu.)}
\thanks{S.\ Gopalakrishnan was with the Department of Electrical and Computer Engineering, University of California, Santa Barbara, CA 93106. He is now with Qualcomm, San Diego, CA 92121. (Email: soorya197@gmail.com.)}}

\markboth{}%
{}

\maketitle


\begin{abstract}
Can we distinguish between two wireless transmitters sending exactly the same message,
using the same protocol? The opportunity for doing so arises due to subtle nonlinear variations across transmitters, even those made by the same manufacturer. 
Since these effects are difficult to model explicitly, we investigate learning device fingerprints using complex-valued deep neural networks (DNNs) that take as input 
the complex baseband signal at the receiver. We ask whether 
such fingerprints can be made robust to distribution shifts across time and locations due to clock drift and variations in the wireless channel.
In this paper, we point out that, unless proactively discouraged from doing so, DNNs learn these strong confounding features rather than the nonlinear
device-specific characteristics that we seek to learn.
We propose and evaluate strategies, based
on augmentation and estimation, to promote generalization across realizations of these confounding factors, using data from WiFi and ADS-B protocols.  
We conclude that, while DNN training has the advantage of not requiring explicit signal models, significant modeling insights are required to focus the learning on the effects we wish to capture.


\end{abstract}


\section{Introduction} \label{sec:intro}

The proliferation of low-cost wireless devices in the Internet of Things (IoT) presents a significant security challenge for the network designer \cite{meneghello2019iot}. A ``fingerprint'' based on physical layer characteristics, capable of distinguishing between devices that transmit exactly the same message, could therefore serve as an important security tool. Such fingerprinting is possible due to subtle hardware imperfections that occur even in devices made by the same manufacturer \cite{Remley2005}. These can provide information regarding the identity and integrity of an IoT device, and may serve as a valuable supplement to conventional security and authentication mechanisms implemented at higher layers of the networking stack.

Wireless fingerprints are often extracted via protocol-specific processing of the received wireless signal \cite{Kohno2005,brik2008wireless,kennedy2008radio,jana2010fast,arackaparambil2010reliability,strohmeier2015passive,radhakrishnan2015gtid,leonardi2017air,merchant2018deep}. 
In this paper, we ask whether it is possible to develop an approach that is independent of the underlying protocol, leveraging the significant advances in 
purely data-driven deep learning over the past decade.  We explore one-dimensional
convolutional neural networks (CNNs) that operate on the complex-valued baseband signal at the receiver, with the goal of determining
the efficacy of extracting fingerprints which are robust to variations across time and location. 

Our results show that deep learning is a promising tool for wireless fingerprinting, while sounding a cautionary note.
The key message is that the network learns the easiest set of features that it can in order to accomplish the desired task (in our case,
discriminating between transmitters based on the received wireless signal), hence we must be extremely proactive in
promoting robustness across effects that we do not want the network to lock on to, which we term {\it confounding factors.} For instance, we would like the radio frequency (RF) signature for a transmitter to be robust across time and for different wireless channels.  
However, if we employ training data collected over a period of time when the channel and carrier frequency offset (CFO) for a transmitter are relatively constant, the CNN will lock onto these rather than to subtle nonlinear effects. This gives unreasonably excellent accuracy on test data collected over the same time period, but disastrous results for data collected on a different day, when both the channel and the CFO can be different. We show that model-based augmentation strategies can significantly improve robustness to such effects.

Our contributions are summarized below.


\begin{figure*}
\includegraphics[width=0.99\textwidth]{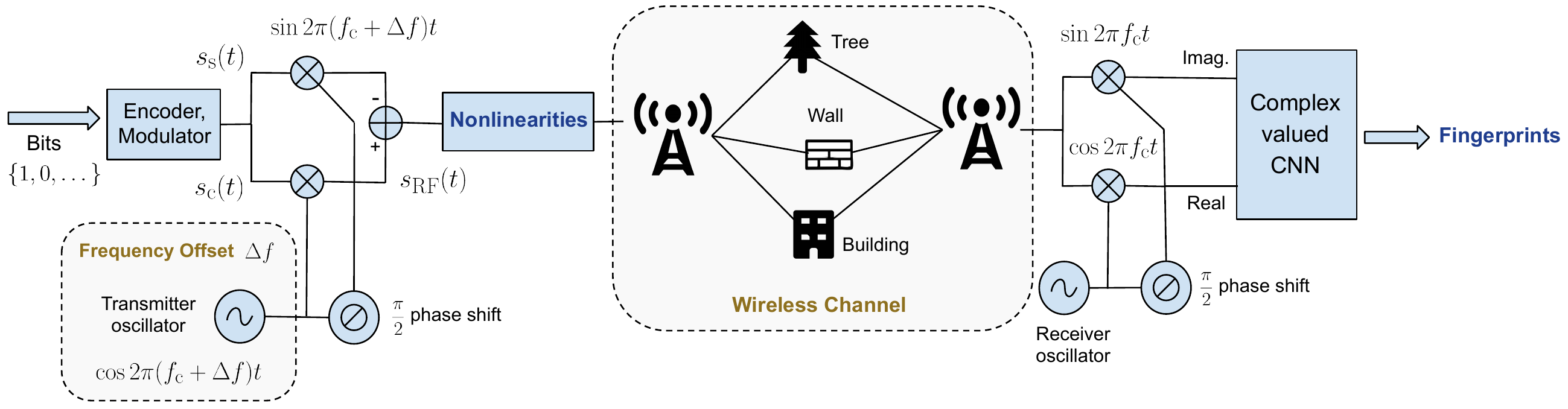}
\caption{Block diagram of a wireless communication system. 
Subtle nonlinearities unique to each device can provide a fingerprint. However, easy-to-learn features such as the CFO and channel are not stable over time and location, affecting generalization.}
\label{fig:comm}
\end{figure*}

\subsection*{Contributions} 

\begin{itemize}[itemsep=2pt, leftmargin=15pt]
\item We demonstrate that protocol-agnostic fingerprinting is possible using complex-valued CNNs, comparing design choices for data from two different wireless protocols: WiFi and ADS-B.
 
\item Using controlled emulations on a clean WiFi dataset, we demonstrate the vulnerability of conventional CNN training to confounding factors such as propagation channels and frequency offsets, which are far stronger than the nonlinear effects we seek to capture. 

\item We develop augmentation strategies based on signal models for the impact of confounding factors, and evaluate performance against compensation techniques that explicitly try to undo them. We find that compensation works well if the undesired features are simple enough, like the CFO. However, for more complex effects such as a multipath channel, model-driven augmentation outperforms explicit estimation and compensation for learning robust signatures.

\item We make publicly available a simulation-based dataset based on models of some typical circuit-level nonlinearities \cite{schenk2008rf,razavi2012rf,wifi1999}. The results we obtain on this dataset are comparable to those from the measurement-based dataset, enabling reproducibility. The dataset and code are available at \cite{repo}.

\end{itemize}


\section{Background and Related Work}

A generic model for a radio frequency (RF) wireless transmitted signal (shown in Fig. \ref{fig:comm}) is as follows:
\begin{equation*}
\srf(t) = \sc(t) \cos 2 \pi \fc t - \ss(t) \sin 2 \pi \fc t
\end{equation*}
where $\fc$ denotes the \textit{carrier} frequency, or the frequency of the electromagnetic wave that ``carries'' 
the information-bearing waveforms $\sc$ (riding on the cosine of the carrier) and $\ss$ (riding on the sine of the carrier). 
Typical parameters for WiFi, for example, are $\fc$ of 2.4 or 5.8 GHz, and $\sc$, $\ss$ having bandwidths of 20 MHz.

The receiver strips the carrier away to recover $\sc(t)$ and $\ss(t)$, and then processes them to decode the information bits
that they carry.  For a typical wireless channel, there are multiple paths from transmitter to receiver, 
so multiple delayed, attenuated and phase-shifted versions of the transmitted waveform sum up at the receiver.
These transformations are best modeled by thinking of the information-bearing waveform as a complex-valued signal,
$s(t) = \sc(t) + j \ss(t)$, where $j = \sqrt{-1}$. The effect of a wireless channel is then modeled as a complex-valued convolution.  The carrier frequency
used at the receiver is not precisely the same as at the transmitter, and the impact of such carrier frequency offset is also most conveniently
modeled in the complex domain.

\subsection{Transmitter-characteristic nonlinearities} \label{sec:background_nonlinearities}

\begin{figure*}[t]
  \centering
  \hfill
  \begin{subfigure}[b]{0.28\textwidth}
    \includegraphics[width=\linewidth]{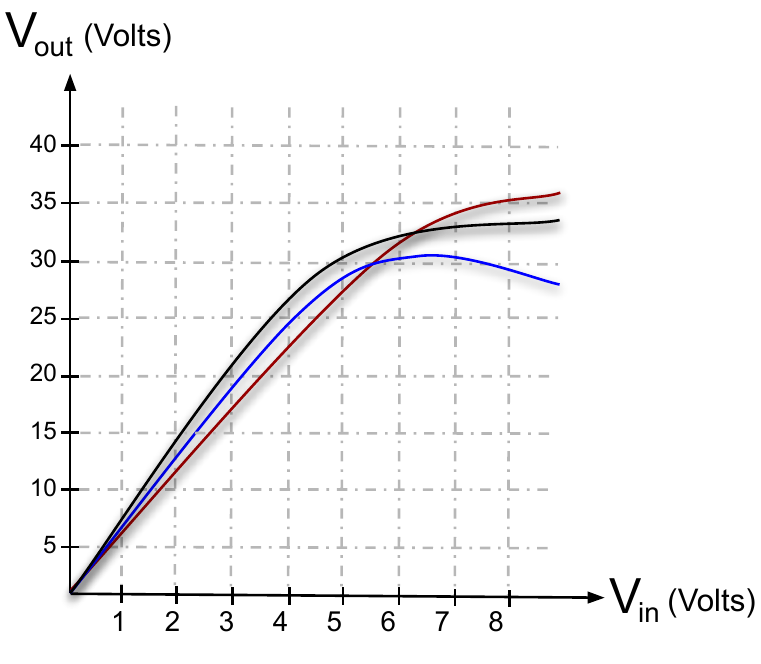}
    \caption{}
     \label{fig:pa}
  \end{subfigure}
  \hfill
  \begin{subfigure}[b]{0.28\textwidth}
    \includegraphics[width=\linewidth]{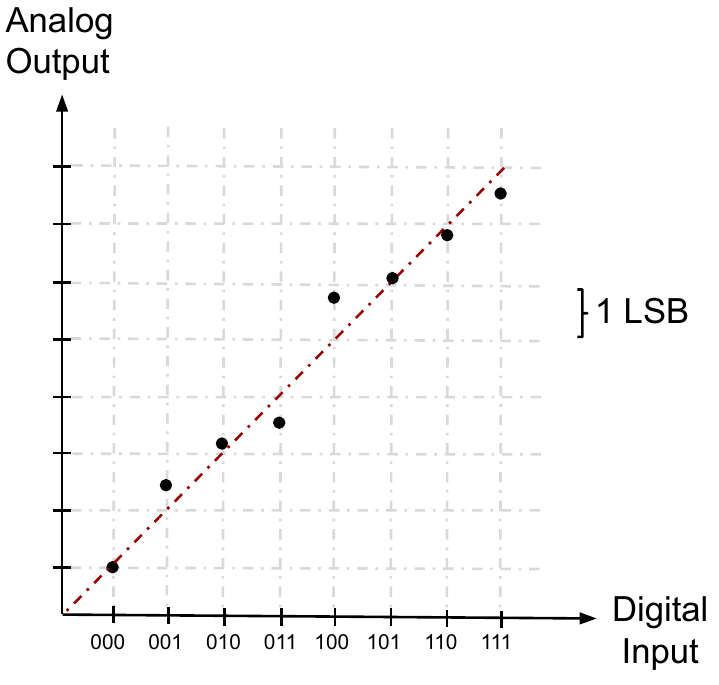}
    \caption{}
 \label{fig:dac}
  \end{subfigure}
  \hfill
  \begin{subfigure}[b]{0.28\textwidth}
    \includegraphics[width=0.8\linewidth]{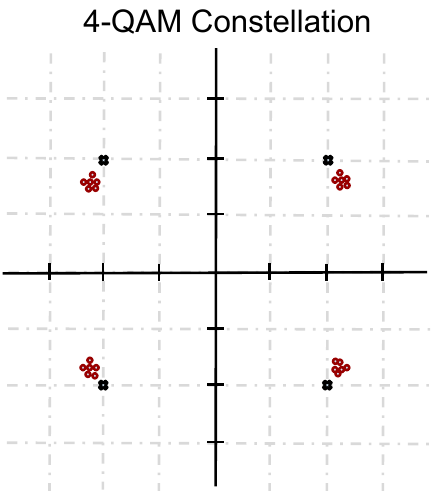}
    \caption{}
     \label{fig:pa}
  \end{subfigure}
  \hfill
  \caption{(a) Example variations of PA nonlinearities across transmitters, (b) Differential nonlinearity caused by DAC, (c) Scatterplots of noisy 4-QAM constellation points with 
  and without I-Q imbalance.}
  \label{fig:nonlinearities}
\end{figure*}

While RF processing is designed to produce as little distortion as possible, in practice, there
are nonlinearities, typically with some characteristics unique to each transmitter because of manufacturing variations, which can in principle provide
RF signatures. 
Variations in components such as digital-to-analog converters (DACs) and power amplifiers (PAs) are inevitable even for transmitters manufactured using exactly the same process. Transistors, resistors, inductors, and capacitors within a device vary around nominal values, typically within a designed level of tolerance, and the goal is to translate the resulting variations in transmitter characteristics into a device signature. We discuss here some example effects, depicted in Figure \ref{fig:nonlinearities}, 
that may contribute towards such a signature.


\begin{itemize}[leftmargin=10pt, itemsep=5pt, topsep=5pt]
\item {\it I-Q Imbalance:} This results from mismatch in the gain and phase of the in-phase (I) and quadrature (Q) signal paths for upconversion \cite{schenk2008rf}. 
The phase of the cosine and sine of the carriers may not be offset by exactly $\pi/2$, and the path gains along the branches may not be equal.
\item {\it Differential Nonlinearity (DNL) due to DAC:} DNL is defined as the discrepancy between the ideal and obtained analog values of two adjacent digital codes
due to circuit component non-idealities \cite{Kadaba1986MOS}. 
\item {\it PA Nonlinearity:} Power amplifiers are ideally linear, but start saturating at high input voltages. There is a significant
literature on PA modeling \cite{Saleh81, Zhu2004Volterra, Ku2003behavioral, Pedro2005overview}, as well as on the impact of PA nonlinearities
on communication systems with high dynamic range such as OFDM \cite{Elena99impact,Merchan98ofdm}.  A common model is a memoryless
polynomial fit (typically up to third order) of the form:
\begin{equation*}
y(t) = a_1x(t) + a_2x^2(t) + a_3x^3(t) + ... + a_nx^n(t)
\end{equation*}
Recent promising results on wireless fingerprints for PA nonlinearities, extracted using CNNs, are reported in \cite{Hanna2019}. 
\end{itemize}

The carrier frequency offset, caused by frequency mismatch in the crystal oscillators at the transmitter and receiver, could also potentially be used as a feature to fingerprint devices \cite{brik2008wireless,leonardi2017air}. However, we treat it here as a confounding factor for our goal of obtaining a fingerprint which is stable over time.
Oscillator frequencies are affected by a few parts per million (ppm) for every 1$^\circ$C change in temperature \cite{razavi2008fundamentals}, and therefore drift daily,
and are also affected by aging \cite{zhou2008frequency}.  The CFO can also be spoofed by a sophisticated enough
adversary manipulating baseband signals \cite{edman2009active,danev2010attacks,merchant2018deep}. While the CFO could still be a useful feature as a defense against simpler attacks (e.g., for systems with relatively frequent transmissions, its slow drift could be tracked across packets to detect abrupt transitions), its role as a confounding factor in our study enables us to benchmark augmentation against
compensation for an effect which can be accurately modeled.

Our goal in this paper, therefore, is to investigate the use of DNNs that extract signatures based on a combination of characteristics such as those in Figure \ref{fig:nonlinearities}, 
treating the CFO and channel as confounding factors to be marginalized over.  For our numerical results, we do not need to explicitly model these nonlinearities, 
since we {\it emulate} the impact of confounding factors on measured data that includes the effect of these nonlinearities, but purely {\it simulated} data based on the models we have developed \cite{repo} yield similar results.

\subsection{Device fingerprinting}

Fingerprints can be extracted from either the transient (microsecond-length) signals transmitted during the on/off operation of devices, or via the steady-state packet information present in between the start and end transients \cite{danev2012physical}. We focus here on work that employs the steady-state method since it is of more practical utility \cite{kennedy2008radio}.
Such prior work can be divided into two categories: (i) approaches that use handcrafted features, and (ii) machine learning based techniques.

\vspace{6pt}
\noindent \textbf{Traditional approaches:} An early approach to device fingerprinting was in \cite{Kohno2005}, albeit only for wired devices in wide area networks. The feature used in \cite{Kohno2005} was the clock skew, which was  observed to be fairly consistent over time, but varied significantly across devices.
This technique was extended in \cite{jana2010fast} to wireless local area networks where timestamps in IEEE 802.11 frames contain more precise information about the clock skew. However, \cite{arackaparambil2010reliability} demonstrated deficiencies of the previous two studies, presenting a spoofing attack based on the clock-skew information generated by a fake access point. 
{In \cite{suski2008using}, WiFi fingerprinting was accomplished by computing the power spectral density of the preamble, followed by cross-correlation to match the spectra of an unknown signal against a bank of known reference spectra. For RFID tags, fingerprinting has been accomplished using power response and timing features for UHF RFID \cite{periaswamy2010fingerprinting,periaswamy2010fingerprinting2,zanetti2010physical}, and a mixture of timing and spectral features for HD RFID \cite{danev2009physical}.}

\vspace{6pt}
\noindent \textbf{Machine learning based approaches:}
There are many papers over the past decade using machine learning to derive fingerprints.
Much of this work involves significant protocol-specific preprocessing, in contrast to the protocol-agnostic approach considered in this paper. 
An early example is the use of support vector machine (SVM) in \cite{brik2008wireless} based on demodulation error metrics such as frequency offset and I/Q offset. {However, this detection method was defeated in \cite{edman2009active,danev2010attacks}, who showed that these modulation features could be impersonated 
via software-defined radios. Other examples of machine learning based, protocol-specific fingerprints include: a $k$-nearest neighbor ($k$-NN) classifier in \cite{kennedy2008radio}
based on spectral analysis of WiFi preambles; linear discriminant analysis (LDA) in  \cite{wang2016wireless} after pilot-aided compensation of RF nonlinearities caused by the receiver; $k$-means clustering of features based on inter-arrival times of ADS-B messages \cite{strohmeier2015passive}; 
a neural network in \cite{radhakrishnan2015gtid} and k-NN in \cite{luo2019transforming} operating on WiFi inter-arrival times; frequency compensation of ZigBee data, followed by a CNN \cite{yu2019robust}; and a CNN operating on the error signal obtained after subtracting out an estimated ideal signal from frequency-corrected received data \cite{merchant2018deep}. Section \ref{sec:spacetime} evaluates the robustness of our approach against protocol-specific estimation strategies, showing that, while estimation works well for simple phenomena such as CFO variations,
the augmentation approach that we study has a clear advantage for more complex effects such as channel variations.
 
Modern CNNs learning directly from I/Q data include \cite{o2016convolutional,o2017introduction} for modulation classification, and \cite{sankhe2018oracle,mcginthy2019groundwork} for device fingerprinting. This line of work employs real-valued networks, with real and imaginary parts of complex data treated as different channels. {Such networks have more degrees of freedom compared to a complex network where the convolution operation is more restricted. Consider a complex convolution operation between input $X$ and weight $W$, resulting in output $Y$:
\begin{equation*}
\re(Y) + j\im(Y) = (\re(W) + j\im(W)) \ast (\re(X) + j\im(X))
\end{equation*}
This can be rewritten in the following form \cite{guberman2016complex,trabelsi2017deep} with the real and imaginary parts of the input stacked as different channels:
\begin{equation} \label{eq:complex_conv}
\begin{bmatrix}
\re(Y) \\ \im(Y) 
\end{bmatrix} 
= 
\begin{bmatrix}
\re(W) & -\im(W) \\ \im(W) & \re(W) 
\end{bmatrix} \ast
\begin{bmatrix}
\re(X) \\ \im(X) 
\end{bmatrix}
\end{equation}
Therefore, a complex network with the CReLU activation function ($\mathrm{ReLU}\left(\re(x)\right) + j\mathrm{ReLU}\left(\im(x)\right)$) can be considered a regularized form of a real ReLU network, with the weight matrix restricted to the structure in (\ref{eq:complex_conv}). This reduction in number of degrees of freedom has been shown to improve generalization performance \cite{hirose2012generalization}. We note that this analysis does not hold for complex networks with the ModReLU activation function ($\mathrm{ReLU}(\left|x\right|) \exp(j\angle x)$), which we find yields better performance than CReLU for our application (Section \ref{sec:arch}); 
ModReLU-based architectures cannot be realized by a real ReLU network. 
It has been observed in recent work that complex networks provide advantages over real networks for the tasks of MRI fingerprinting \cite{virtue2017better}, radar-based terrain classification \cite{zhang2017complex}, audio source separation \cite{lee2017fully}, music transcription \cite{trabelsi2017deep} and channel equalization \cite{scardapane2018complex}. Our results in the appendix on the gain provided for the fingerprinting problem are in line with such prior work, and motivate further exploration of neural networks tailored to complex-valued data.}
{It is worth noting that, for real-valued networks, standard DNNs and CNNs are compared with multi-stage training (MST) of simple building blocks for fingerprinting in \cite{youssef2018machine}, with MST yielding the best performance. Such work highlights the need for continued architectural experimentation for both real-
and complex-valued networks.}  

The present paper builds on our conference paper \cite{globecom2019}, which considers the impact of ID spoofing and SNR on CNN-based fingerprinting. To our knowledge, \cite{globecom2019} was the first to employ complex-valued CNNs for wireless fingerprinting; it precedes and is independent of \cite{agadakos2019deep}, which also uses complex-valued networks.
While a part of the discussion from \cite{globecom2019} is included here in order to provide a complete treatment, the main focus of this paper is different: we investigate robustness of fingerprints to variations in the CFO and wireless channel. While \cite{globecom2019} considers noise augmentation to handle SNR
mismatch between training and test data, in the present paper, we consider
augmentation and compensation strategies for CFO and channel, and introduce the concept of test time augmentation for handling confounding factors. {We should note that the concept of test time augmentation proposed here is different from classical ensemble methods such as boosting or bagging  \cite{freund1995desicion, friedman2001greedy}: rather than averaging over an ensemble of machines, we are averaging over an ensemble of inputs.  Given recent
promising results on the use of boosting techniques in multilayer settings \cite{moghimi2016boosted,feng2018multi,chen2018deep}, it is of interest to explore comparison and possibly combination of such techniques
with our augmentation strategy for deriving RF signatures robust to confounding factors.}


\begin{figure}[b]
\centering
\hspace{-5pt}
\begin{subfigure}{0.49\linewidth}
	\begin{center}
	\scalebox{0.6}{
	\begin{tikzpicture}
		\begin{axis}[
			axis lines = center,
			set layers=standard,
			xmin=-0.5, xmax=0.5, ymin=-0.5, ymax=0.5,
			axis equal,
			xlabel = {\large $\re(x)$},
			ylabel = {\large $\im(x)$},
			x label style= {at ={(axis cs: 0.62, -0.13)}},
			yticklabels={,,},
			xticklabels={,,},
			]
			\begin{pgfonlayer}{axis background}
				\filldraw[fill=white!50, color=blue!6] (axis cs: 0, 0) circle [radius=0.23];
			\end{pgfonlayer}
			\node [fill=blue!6] at (axis cs: -0.005, 0.08) (a) {\large $y=0$};
			\node at (axis cs: 0.258, -0.053) (a) {\large $b$};
			\node at (axis cs: -0.29, -0.053) (a) {\large $-b$};
			\node at (axis cs: -0.28, -0.3) (a) {\large 
				$\begin{aligned} 
					\phase{y} &= \phase{x}    \\[2pt]
					|y| &= |x| - b
				\end{aligned}$
				};
		\end{axis}
	\end{tikzpicture}}
	\end{center}
	\caption{$y=\modrelu(x)$}
	\label{fig:modrelu}
\end{subfigure}
\hfill
\begin{subfigure}{0.49\linewidth}
	\begin{center}
	\scalebox{0.6}{
	\begin{tikzpicture}
	\begin{axis}[
			axis lines = center,
			set layers=standard,
			xmin=-0.5, xmax=0.5, ymin=-0.5, ymax=0.5,
			axis equal,
			xlabel = {\large $\re(x)$},
			ylabel = {\large $\im(x)$},
			x label style= {at ={(axis cs: 0.62, -0.13)}},
			xticklabels={,,},
			yticklabels={,,},
			legend style={at={(-0.2, 0.2)}},
			]
			\begin{pgfonlayer}{axis background}
				\filldraw[fill=white!100, color=blue!6]
								(axis cs: -0.6, -0.5) rectangle (axis cs: 0, 0);
				\filldraw[fill=gray!4, draw=none]
								(axis cs: -0.6, 0) rectangle (axis cs:0 , 0.5);
				\filldraw[fill=gray!6, draw=none]
							(axis cs: 0, -0.5) rectangle (axis cs: 0.6, 0);
			\end{pgfonlayer}
		 	\node at (axis cs: -0.32, -0.27) (a) {\large 
		 		$y=0$
		 		};
			\node at (axis cs: -0.3, 0.22) (a) {\large 
				$y =  j \im(x)$
				};
			\node at (axis cs: 0.3, -0.27) (a) {\large 
				$y = \re(x)$		
				};
			\node at (axis cs: 0.25, 0.22) (a) {\large 
				$y=x$			
				};
	\end{axis} 
	\end{tikzpicture}}
	\end{center}
	\caption{$y=\crelu(x)$}
	\label{fig:crelu}
\end{subfigure}
\vspace{5pt}
\caption{ModReLU and CReLU activation functions in the complex plane. 
ModReLU preserves the phase of all inputs outside a disc of radius $b$, while CReLU distorts all phases outside the first quadrant. 
Figure adapted from \cite{trabelsi2017deep}.}
\label{fig:activation_functions}
\end{figure}
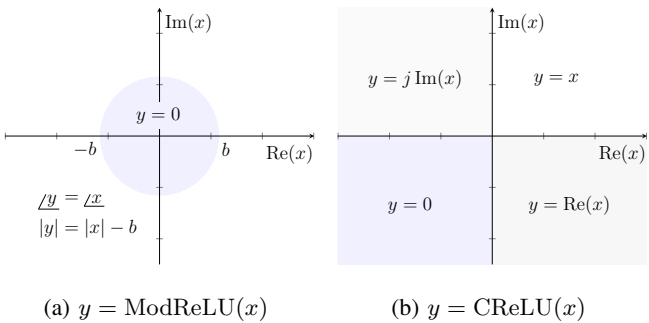

\begin{figure*}[t]
\centering
\includegraphics[width=0.98\textwidth]{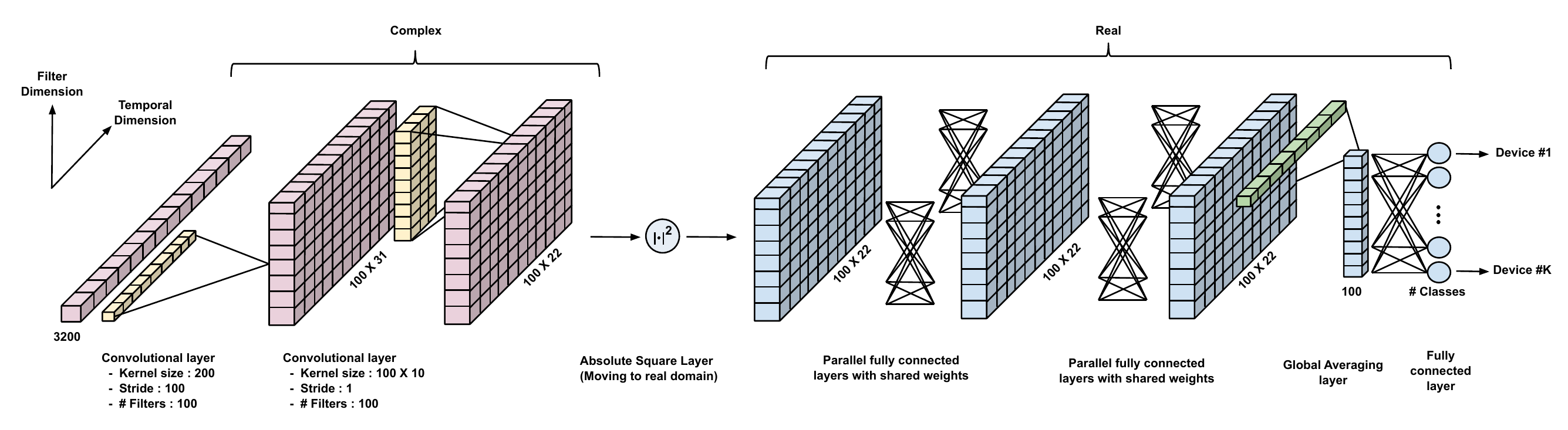}
\caption{Complex-valued 1D CNN architecture for WiFi signals.}
\label{fig:arch}
\vskip -0.1in
\end{figure*}

In  \cite{restuccia2019deepradioid}, channel-resilient fingerprinting was studied by modifying the transmitter using a finite impulse response (FIR) filter. Our work on channel resilience is based solely on modifying DNN training and does not involve transmitter-side alterations. {In recent work, \cite{al2020exposing,jian2020deep} reported a significant degradation in accuracies when training and test data were from different days, with fingerprints extracted using real-valued CNNs. While equalization was observed to improve performance in the different day scenario, it caused a drop in accuracy when training and test data were from the same day. These results are in line with our observations in Section \ref{sec:channel}: while equalization can help, the residual error from this approach appears to swamp out the nonlinear characteristics we are interested in. We find model-based augmentation to be a more effective strategy for learning robust fingerprints.}


\section{Complex-valued Representations} \label{sec:arch}

The subtle nonlinear effects discussed in the previous section are difficult to model explicitly, hence deep learning is a natural approach to teasing out transceiver signatures based on them.
We explore the use of complex-valued neural networks for this purpose: these are well-matched to the complex baseband received signal. Such networks have previously been used for speech, music and vision tasks \cite{wisdom2016full,trabelsi2017deep}. Here, we learn device fingerprints for two different wireless protocols: WiFi and ADS-B.

\begin{figure}[b]
\centering
\includegraphics[width=0.95\columnwidth]{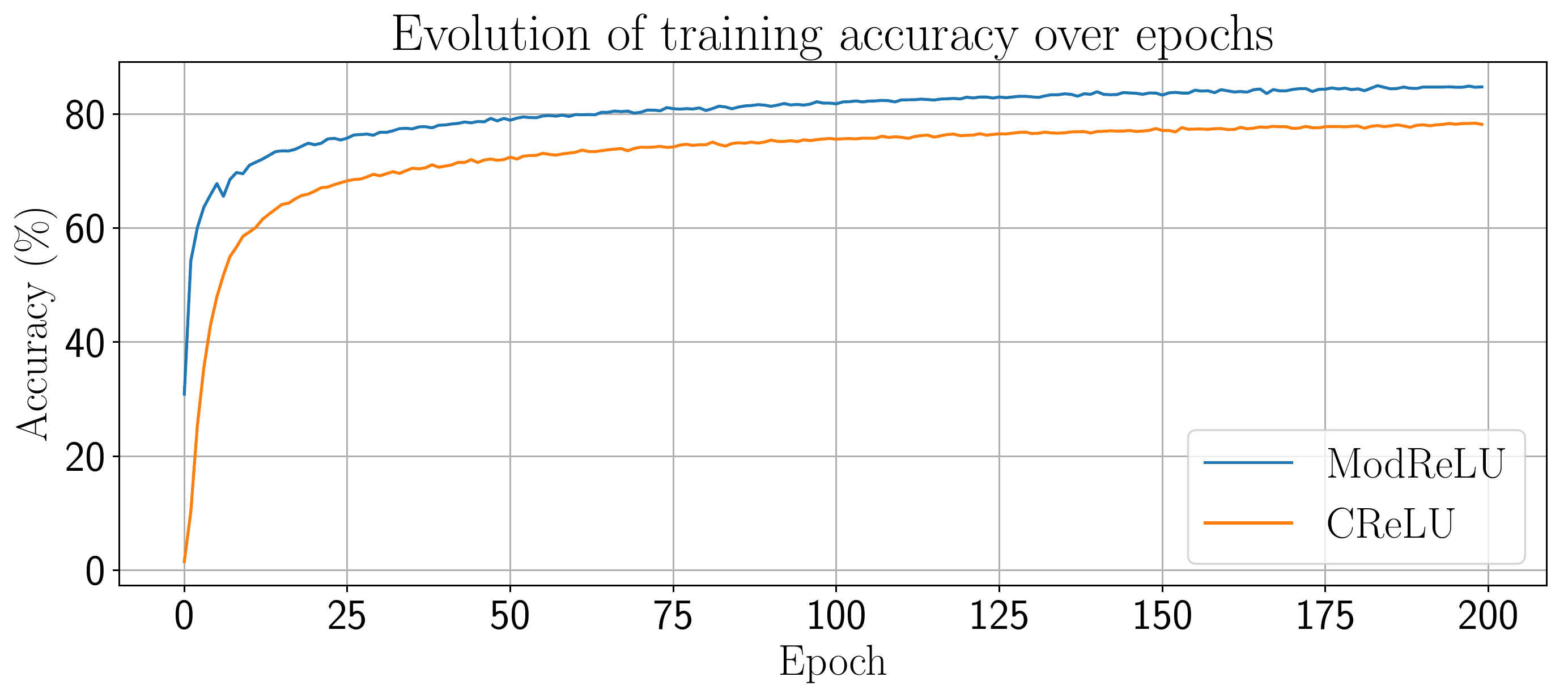}
\caption{Evolution of training accuracy over epochs for ModReLU and CReLU networks (ADS-B). ModReLU provides a small gain in train and test accuracies over CReLU, with similar convergence behavior.
}
\label{fig_train_test_acc}
\end{figure}

\vspace{6pt}
\noindent \textbf{Data:} We provide results for the following external database: 
\begin{itemize}[itemsep=3pt, topsep=2pt]
\item WiFi data containing a mix of IEEE 802.11a ($\fc = 5.8\text{ GHz}$) and IEEE 802.11g ($\fc = 2.4 \text{ GHz}$) packets from 19 commercial-off-the-shelf devices, collected indoors without channel distortion using a Tektronix RSA5126B receiver.
\item ADS-B air traffic control signals ($\fc = 1.09 \text{ GHz}$, narrowband) collected in the wild from 100 airplanes over a span of 10 days, using a Tektronix RSA5106B receiver. These signals are used for transmitting airplane position and velocity information to ground stations.
\end{itemize}
We use available oversampled data for both protocols, with WiFi signals sampled at 200 MHz and ADS-B at 20 MHz. The length of the preamble is then 3200 samples for WiFi and 320 samples for ADS-B.

\vspace{6pt}
\noindent \textbf{Architecture:} For complex layers, we explore the following choices of activation functions, shown in Figure \ref{fig:activation_functions}:

\begin{itemize}[itemsep=2pt, topsep=5pt]
\item \textit{ModReLU} - This function affects only the magnitude and preserves phase. Here $b$ is a learned bias. 
\begin{equation*}
\modrelu(x) = \max\left(|x| - b, 0\right) \,\,e^{j\phase{x}}.
\end{equation*}
\item \textit{CReLU} - Here, separate ReLUs are applied to the real and imaginary parts of the input. The phase of the output is therefore restricted to $[0, \pi/2]$.
\begin{equation*}
\crelu(x) = \max\left(\re(x), 0\right) + j\max\left(\im(x), 0\right).
\end{equation*}
The loss in phase information can be potentially compensated by using wider filters (i.e.\ with a larger number of channels) capable of providing phase derotation. 
\end{itemize}

Figure \ref{fig:arch} depicts the complex-valued 1D CNN  we use for WiFi signals, using as input the I/Q data at the receiver, restricted to the preamble. $\text{An}\; |\cdot|^2\; \text{layer}$ is used midway through the network to convert complex representations to real ones. The network architectures we use are listed below in compact form (similar to the notation in \cite{ciresan2012multi}):

\begin{itemize}[itemsep=5pt, topsep=5pt]
\item \textit{ADS-B}: $100 \, C \, 40 \times 20 -100 \, C \, 5 \times 1 - |\cdot|^2 - \mathrm{\textit{Avg}} - 100\, D$.
\item \textit{WiFi}: $100 \, C \, 200 \times 100 - 100 \, C \, 10 \times 1 - |\cdot|^2 - 100 \, D - 100 \, D - \mathrm{\textit{Avg}}$.
\end{itemize} 

\noindent The notation should be read as follows: 
\begin{itemize}[itemsep=2pt, topsep=3pt]
\item $\left< \text{number of filters}\right>\; C \; \left< \text{convolution size}\right> \times \left< \text{stride} \right>$
\item $\left< \text{number of neurons} \right> \; D$
\end{itemize}
where $C$ denotes a convolutional layer, $D$ a fully connected layer, and \textit{Avg} a temporal averaging layer.

Complex backpropagation is performed using the framework of \cite{trabelsi2017deep}, taking partial derivatives of the cost with respect to the real and imaginary parts of each parameter. We use 200 samples per device for training and 100 for testing for WiFi, and 400 samples per device for both training and testing for ADS-B. Detailed information about hyperparameter choices, cross-validation, etc.\ is provided in the appendix.
Code is available at \cite{repo}.

\vspace{5pt}
\noindent \textbf{Performance:} 
Using the preamble alone, we obtain 99.62\% fingerprinting accuracy for 19 WiFi devices, and 81.66\% accuracy for 100 airplanes using the ADS-B protocol. We find that the ModReLU architecture outperforms CReLU (shown in Fig. \ref{fig_train_test_acc}), without any difference in convergence speed. 
The appendix provides a performance comparison to real-valued CNNs, along with a visualization of input signals that strongly activate filters in the trained CNNs.

\vspace{-0.1pt}


\section{Stability to CFO and Channel Variations} \label{sec:spacetime}

In this section, we use the clean WiFi dataset for controlled experiments emulating the effect of frequency drift and channel variations. We show that these fluctuations can have a disastrous effect on performance and study compensation and augmentation strategies to promote robustness. 

\subsection{Nuisance Parameters, Compensation and Augmentation} \label{sec:augmentation}

Before providing specific results, we lay out our overall framework.

Consider input data $\bx$ (the packet preamble in our case) fed to a neural network which aims to classify the device ID $y$.  In our present context, we may think of this input data as a transformation
of an ideal input $\bxideal$ capturing the desired characteristics of the device, passed through a transformation $f_{\theta}$, where $\theta$ is a nuisance parameter such as
a CFO or channel: $\bx = f_{\theta} ( \bxideal )$.  A network trained with such inputs would ideally produce posteriors $p(y| \bx) = p(y|f_{\theta} ( \bxideal ))$ as the softmax
outputs.  In the scenarios of interest, we define a single ``day'' of training as a scenario in which $\theta$ is fixed during the training period for a given device, but differs across
different devices.  In this case, it is natural for the DNN to use information in $\theta$ to classify devices.  Indeed, if the discrimination based on $\theta$ is easier than that
based on the subtle nonlinear signatures buried in $\bxideal$, then the DNN will focus on using $\theta$ rather than the information in $\bxideal$.  When we then
test on a different ``day'' when the value of the nuisance parameter $\theta$ is different, we understandably get poor performance.

\vspace{5pt}
\noindent {\bf Compensation:} If we have detailed protocol-level information and good enough models, then it is possible to try to invert $f_{\theta}$ to recover $\bxideal$ from $\bx$, and to then train the DNN based on
this estimate.  For example, we can estimate and undo a CFO, or equalize a channel.  For the particular experiments we do, we find that compensation works well for simple nuisance parameters such as the CFO, but that the
residual errors after equalization are enough to swamp out the subtle nonlinear effects we are after.  

\vspace{5pt}
\noindent {\bf Augmentation:} An alternative to protocol-specific compensation strategies is to use models for how the nuisance parameters operate on the input to augment the data.
Specifically, we create new inputs of the form $\bx' = f_{\thetaaug} ( \bx )$, where we choose $\thetaaug$ from a set $\Theta$ such that
\begin{equation*}
\bx' = f_{\thetaaug} ( \bx ) =f_{\thetaaug} \left( f_{\theta} ( \bxideal ) \right) \approx f_{\theta'} (\bxideal) ~,~ \theta' \in \Theta
\end{equation*}
where $\theta'$ is an ``effective'' nuisance parameter.  Now, if we train the DNN using multiple augmentations of $\bx$, then we hope that the network learns to use $\bxideal$ to a greater extent than before, since we are varying $\theta'$ for a given device. Nevertheless, standard training does not {\it guarantee} marginalization over $\theta'$. Rather, it allows the network to produce posteriors of the form 
$p(y| \bx') = p \left( y|f_{\thetaaug} \left( f_{\theta} ( \bxideal ) \right) \right) \approx p \left( y|f_{\theta'} ( \bxideal  )\right)$, where hopefully the information from $\bxideal$ is being used to a greater extent because of training
augmentation.  When we are now presented with a fresh test input $\bx = f_{\theta} ( \bxideal)$, we are not
guaranteed that this particular realization of the nuisance parameter $\theta$ is comfortably far from the decision boundaries that the network has learnt.  On the other hand,
test time augmentation allows us to generate multiple effective nuisance parameter realizations which we can average over.
\begin{equation} \label{eq:testaug}
\frac{1}{| \Theta_{\mathrm{test}} |} ~ \sum_{\thetaaug \in \Theta_{\mathrm{test}}} p \left(y| f_{\thetaaug} \left( f_{\theta} ( \bxideal ) \right) \right)
\end{equation}
Thus, we are effectively averaging over $| \Theta_{\mathrm{test}} |$ realizations of the ``effective'' nuisance parameters $\theta'$.

\vspace{5pt}
\noindent {\bf Residual approach:} An interesting way to combine the above two strategies is by excising a reconstruction of the transmitted message based on a linear model to obtain a residual signal containing device nonlinearities. Using the known preamble sequence and estimated CFO and channel, we can compute an ideal noiseless reconstruction $\bxhat$ of the received signal $\bx$. The residual noise $\bx-\,\bxhat$ can then fed as input to a neural network. Since this residual signal still contains CFO and channel effects, we find that this technique does not work well on its own. However, it can be used in combination with augmentation to confer robustness. 

In the following sections, we assess performance using the average of five different runs, with different random realizations of CFOs and channels used for emulation and augmentation, as well as different random seeds for CNN weight initialization. In all graphs, error bars denote one standard deviation from the mean over different runs.


\subsection{Carrier Frequency Offset} \label{sec:cfo}

We first examine robustness to the carrier frequency offset (CFO), which we treat as a confounding factor due to its drift over time and vulnerability to spoofing (Section \ref{sec:background_nonlinearities}). 
We investigate this by inserting offsets in data, emulating an oscillator frequency tolerance of $\pm \,20$ parts per million as specified in the IEEE 802.11 standard \cite{wifi1999}. We begin with an example where only the test data is offset.

\vspace{5pt}
\noindent\textbf{Offset in test data alone:} We find that networks trained on clean data do \textit{not} generalize to offset data, even when the offset is very small: as shown in the first row of Table \ref{table_offset_test}, accuracy drops to 4.6\% at an offset of 20 ppm. 
In order to alleviate this, we augment the training set with randomly chosen CFOs and report results in the second and third rows of Table \ref{table_offset_test}. We consider two types of random offsets: $\text{Bernoulli}\,\{-20, 20\}$ ppm and $\text{uniform}\,(-20, 20)$ ppm, augmenting the size of the training set by 5x in each scenario.

\begin{table}[t]
\caption{Performance when only the test data is offset, with CFOs in the range (-20, 20) ppm. {The first row shows that this results in poor accuracies if we do not modify our training strategy. Rows 2 and 3 then demonstrate that augmenting training data with uniformly distributed CFOs helps confer robustness.}
 }
\label{table_offset_test}
\vskip 0.05in
\centering
\begin{small}
\setlength\extrarowheight{1pt}
\begin{tabularx}{0.9\columnwidth}{lXXX}%
	\toprule
	\multirow{2}{*}{\begin{tabularx}{1.8cm}{@{}l}\\[-11pt]Type of data \\[-2pt] augmentation\end{tabularx}} & \multicolumn{3}{c}{CFO in test set} \\ 
	\cmidrule(l){2-4} & None & Bernoulli  & Uniform \\
	\midrule
	\\[-13pt]
	None & 99.50 & 4.63 & 13.58 \\[1pt]
	Bernoulli & 3.32 & 99.32  & 13.53\\[1pt]
	\rowcolor{LightCyan} Uniform & 96.21 & 90.79  & 95.37 \\[0.2pt]
	\bottomrule
\end{tabularx}
\end{small}
\vskip -0.05in
\end{table}

\begin{table}[b]
\caption{Effect of augmentation in the ``different day'' CFO setting, with CFOs in the range (-40, 40) ppm. ``Random'' training augmentation uses a new randomly chosen CFO for each packet, while the ``orthogonal'' type uses the same set of offsets across devices. In both cases, the offsets are drawn from a uniform distribution. 
 }
\label{table_offset_dd}
\vskip 0.05in
\centering
\begin{small}
\setlength\extrarowheight{1pt}
\begin{tabularx}{0.98\columnwidth}{l@{}c XXXX}%
	\toprule
	\multicolumn{2}{@{\hspace{1pt}}l}{
	  	\multirow{2}{*}{
	  		\begin{tabularx}{2.2cm}{@{\hspace{2.5pt}}l}
	  			\\[-11pt]Training \\[-2pt] augmentation
	  		\end{tabularx}
	  		}
		} 
		& \multicolumn{4}{c}{Test time augmentation}\\[1pt]
	 	\cmidrule(l){3-6}
		& & None & 5 & 20 & 100\\[1pt]
	\midrule
	\\[-12pt]
	None  & -- & 9.68 & 7.84 & 8.74 &  8.47\\[1pt]
	Random  & 5 & 74.21 & 71.84 & 74.21 & 77.37 \\[1pt]
	 & 20  & 72.79 & 75.84 & 78.05 & 80.05 \\[1pt]
	Orthogonal & 5 & 69.58 & 75.11 & 81.05 & 83.63
	\\[1pt]
	 & 20  & 82.37 & 82.32 & 86.21 & 87.11\cellcolor{LightCyan} \\
	\bottomrule
\end{tabularx}
\end{small}
\end{table}

This strategy can significantly help in learning robust fingerprints, but the type of augmentation matters: in particular, it is insufficient to augment with worst-case offsets alone. When we train with Bernoulli offsets, the network becomes robust to Bernoulli test offsets (99.3\%), but fails to generalize to any offset smaller than 20 ppm, including an offset of zero. In contrast, when we augment data with uniformly chosen offsets, we obtain resilience (>90\%) to all test set offsets in the desired range.

\vspace{5pt}
\noindent \textbf{"Different day" scenario (no augmentation or compensation):} We now emulate collecting training data on one day and testing on another: 
{given clean data $\bxideal$, we add CFOs $\theta$ to emulate the effect of different days: $f_{\theta}(\bxideal)$}. We insert different ``physical'' offsets for each device, but fix the offset for all packets from a particular device. The offsets are randomly chosen in the range $(-40, 40)$ ppm (since both the transmitter and receiver oscillators can vary by $\pm 20$ ppm). Oscillator drift across days is realized via different random seeds for training and test offsets. 

This ``different day'' setting makes it particularly easy for the network to focus on the CFO as a fingerprint: {since each device has a different offset on each day, training on a single day leads to the DNN focusing on using the CFO as a means of distinguishing between devices}. This results in artificially high training accuracies (94.2\%), but poor test set performance (9.7\%) on a different day when the devices have different CFOs. We now explore two strategies to restore performance: data augmentation with randomly chosen CFOs, and frequency compensation.

\vspace{5pt}
\noindent\textbf{"Different day" scenario with augmentation:}
{In order to promote robustness, we add new, randomly chosen CFOs $\thetaaug$ \textit{on top} of the CFOs used for different day emulation: $f_{\thetaaug}(f_{\theta}(\bxideal))$.
Table \ref{table_offset_dd} reports on the efficacy of various CFO augmentation strategies, capable of increasing test accuracy to 87.1\%. 
For training data, we find that the best augmentation technique is to use a different augmentation offset for each packet from a device, but the same set of offsets across devices, which discourages
the network from learning the CFO as a means of distinguishing between devices.  We term this an ``orthogonal'' strategy: we are trying to train in a direction ``orthogonal'' 
to the tendency to lock onto the ``physical'' CFO as a signature.


\begin{table}[b]
\caption{Comparison of augmentation, compensation and the residual approach in the ``different day'' CFO scenario. 
 The training and test datasets are augmented by 20 and 100 times respectively.}
\label{table_cfo_baseline}
\vskip 0.05in
\centering
\begin{small}
\setlength\extrarowheight{1pt}
\begin{tabularx}{0.98\columnwidth}{lXX}%
	\toprule
	 Training strategy 
	 & Test accuracy \\[1pt]
	\midrule
	\\[-12pt]
	Baseline (no augmentation or compensation) & 9.68 \\[1pt]
	Augmentation  & 91.47   \\[1pt]
	Residual + Augmentation & 93.21   \\
	Compensation   & 96.37  \\[1pt]
	\bottomrule
\end{tabularx}
\end{small}
\end{table}

A novel finding is that {\it data augmentation for testing} leads to significant performance gains when we add up soft outputs across augmented versions of each test packet. The best result is obtained when we insert a different randomly chosen CFO for each of a 100 copies of each test data packet, and then sum up the softmax outputs across the augmented data. 
We find that averaging of logits also improves performance, but not to the extent of the softmax average.

\vspace{5pt}
\noindent\textbf{"Different day" scenario with frequency compensation:} We can also estimate and correct the offset using knowledge of the periodic structure of the preamble.
Consider a periodic signal $s[n]$ with period $L$, and frequency offset $\theta$ resulting in $r[n] = s[n] \,\exp(j2\pi n \, \theta)$.
Since we know that $s[n]=s[n+L]$, the CFO can be estimated by correlating $r$ with its shifted version:
\begin{align*}
\widehat{\theta} = \frac{1}{2\pi L}\,\angle{\bigg(\sum_n r[n] \,r^*[n+L]\bigg)}.
\end{align*}
We follow a two-step approach \cite{sourour2004frequency} involving a coarse estimate from the 802.11 short training sequence ($L=16$) and then a fine estimate from the long training field ($L=64$).
This method restores accuracy to 96.4\%, and, as shown in table \ref{table_cfo_baseline}, its accuracy is about 4.9\% better than that with augmentation. 

\vspace{5pt}
\noindent\textbf{Residual approach:} 
We could also use the estimated CFO to compute a residual signal that can be fed as input to a CNN, as described in Section \ref{sec:augmentation}. 
This approach can be combined with augmentation to obtain a performance improvement over pure augmentation, as shown in Table \ref{table_cfo_baseline}.  Stripping out the message in this manner makes it easier for the network to learn nonlinear signatures.


\begin{figure*}[t]
  \centering
  \begin{subfigure}[b]{0.31\textwidth}
    \includegraphics[width=\linewidth]{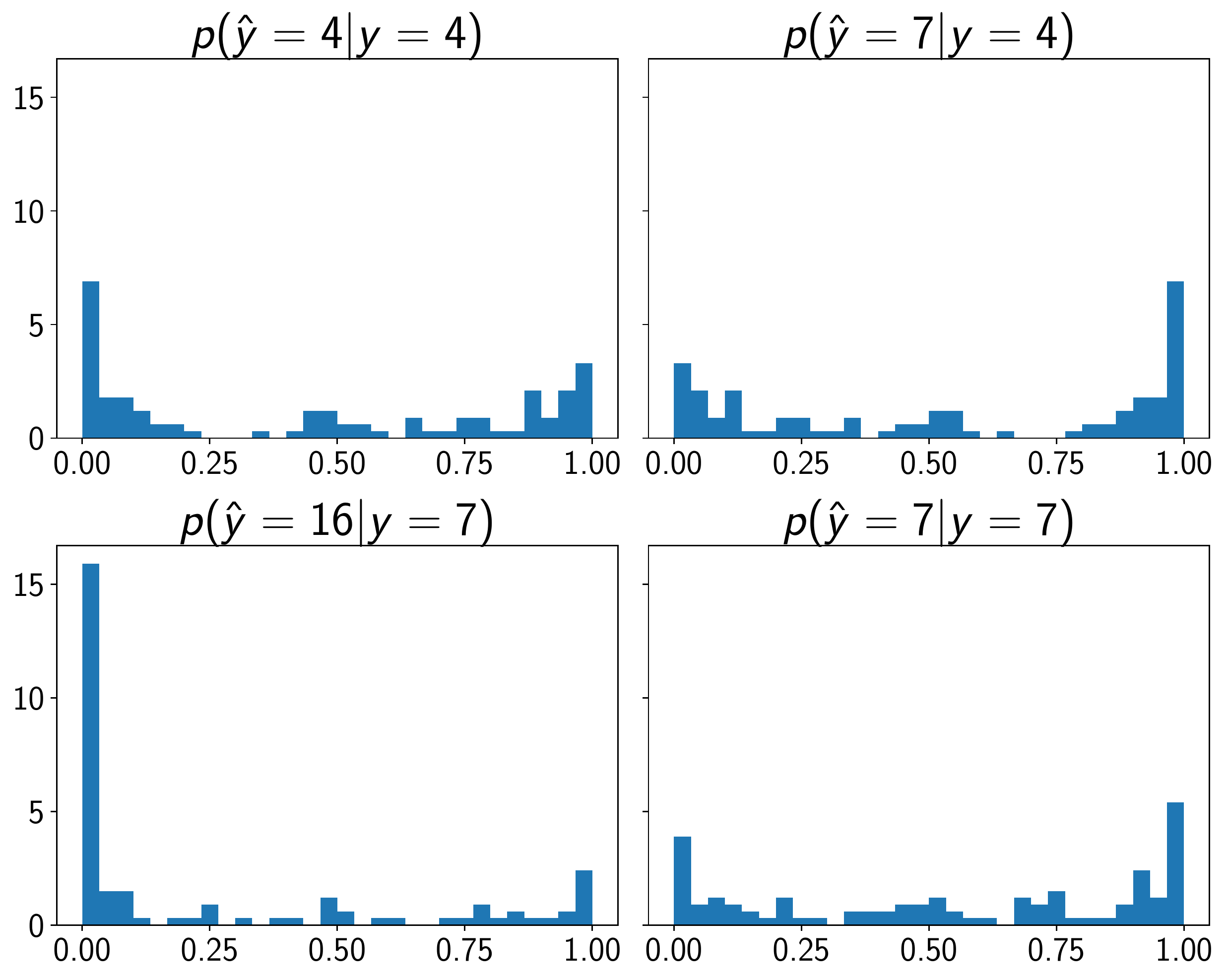}
    \caption{No test augmentation.}
     \label{fig:0test}
  \end{subfigure}
  \hfill
  \begin{subfigure}[b]{0.31\textwidth}
    \includegraphics[width=\linewidth]{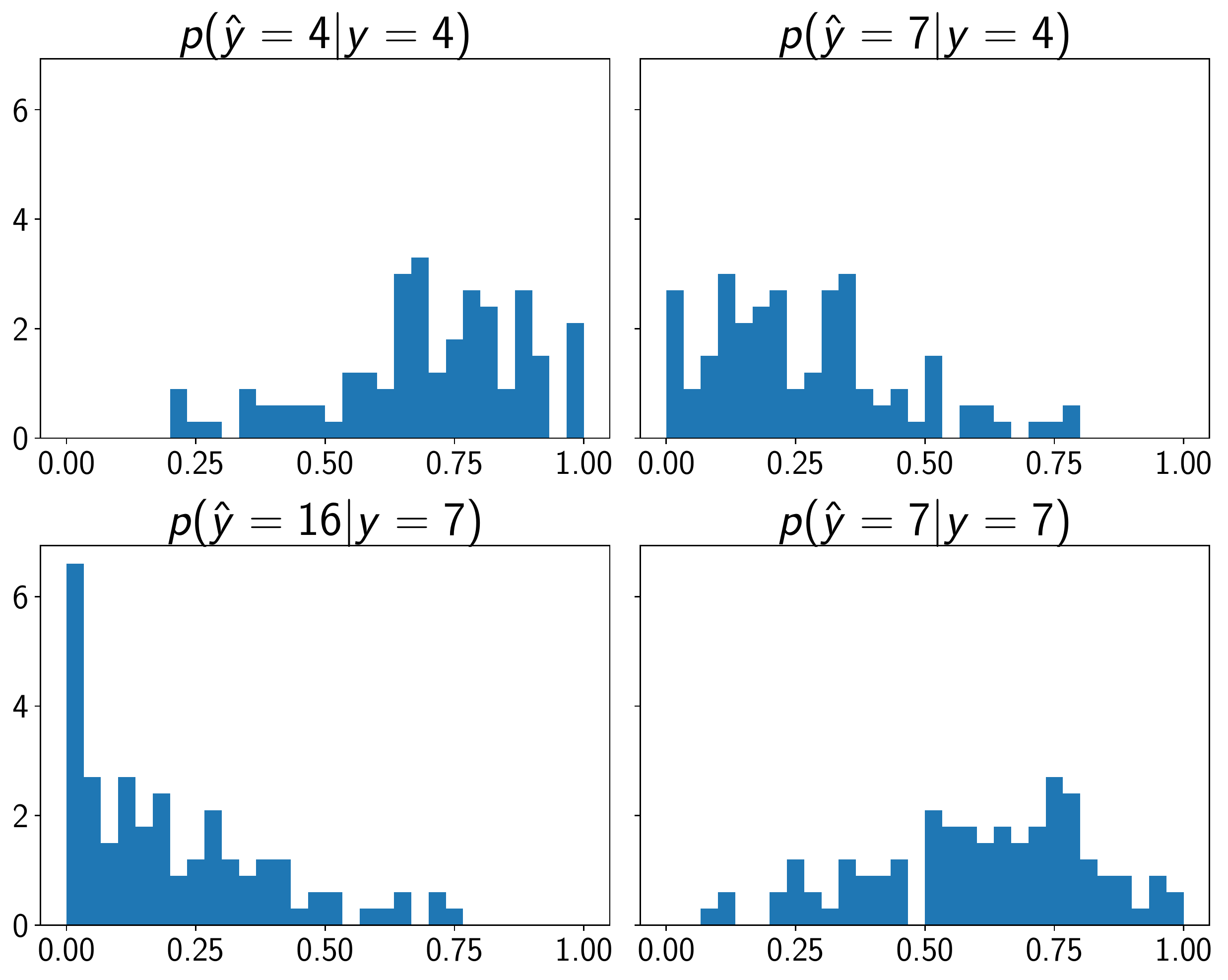}
    \caption{10 test augmentations.}
 \label{fig:10test}
  \end{subfigure}
  \hfill
  \begin{subfigure}[b]{0.31\textwidth}
    \includegraphics[width=\linewidth]{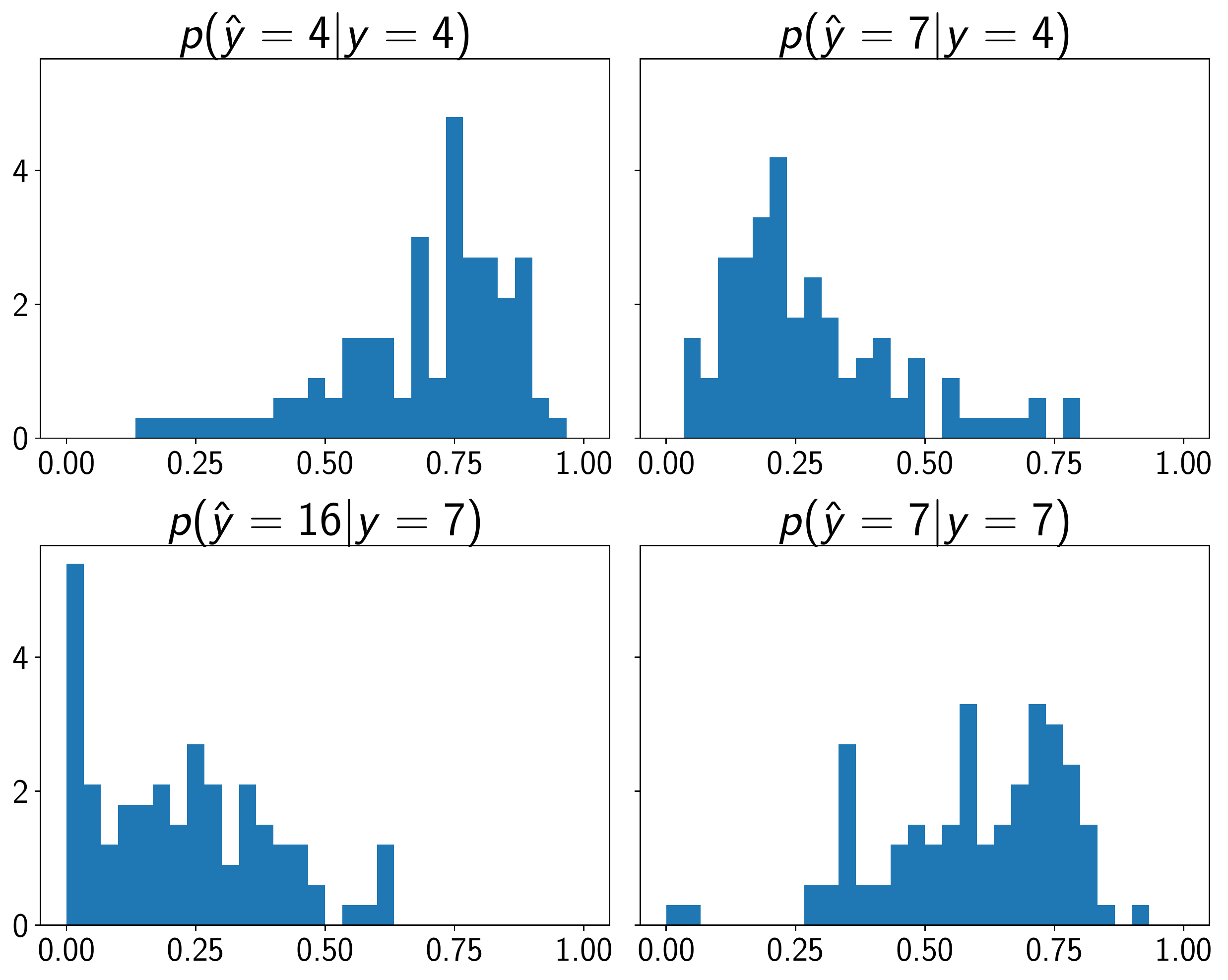}
    \caption{100 test augmentations.}
     \label{fig:100test}
  \end{subfigure}
  \vskip .5em
  \caption{Plots showing how test augmentation affects the histogram of softmax outputs $p(\hat{y})$ (averaged over augmentations) for data from two specific devices ($y=4$ and $y=7$), in the ``different day'' channel setting. Histograms are normalized to be probability densities. As the number of test augmentations increases, the probability of correct prediction $p(\hat{y}=4|y=4)$ and $p(\hat{y}=7|y=7)$ shifts towards $1$. }
  \label{fig:testaughist}
  \vskip -.5em
\end{figure*}


\subsection{Multipath Channels} \label{sec:channel}

The wireless channel is another important source of distribution shift between training and test data. Since multipath components in the channel depend on propagation geometry, a network that locks on to the channel will fail to generalize to test data collected on a different day or location. If the training data does not span a sufficiently diverse set of geometries, it could contain channels that are highly correlated with the transmitter ID, necessitating the use of channel augmentation or equalization strategies to improve robustness. 

\begin{table}[t]
\centering
\caption{Power-delay profile for the EPA multipath fading model. Tap amplitudes $A_k$ are Rayleigh distributed with variance $P_k$.}
\label{table_EPA}
\vskip 0.05in
\setlength\extrarowheight{1pt}
\begin{small}
\begin{tabularx}{0.95\columnwidth}{cccccccc}%
	\toprule
	$k$ & 1 & 2 & 3 & 4 & 5 & 6 & 7\\[1pt]
	\midrule
	$\tau_k$ (ns) & 0 & 30 & 70 & 90 & 110 & 190 & 410 \\[1pt]
	$P_k$ (dB) & 0.0 & -1.0 & -2.0 & -3.0 & -8.0 & -17.2 & -20.8\\[1pt]
	\bottomrule
\end{tabularx}
\end{small}
\end{table}

\begin{table}[!b]
\caption{Performance in the ``different day'' channel setting when we train on 2 days and test on a third day. ``Random'' augmentation uses a randomly drawn channel for each packet, while the ``orthogonal'' type uses the same set of channels across devices. 
 }
\label{table_ch_dd}
\vskip 0.05in
\centering
\begin{small}
\setlength\extrarowheight{1pt}
\begin{tabularx}{0.98\columnwidth}{l@{}c XXXXX}%
	\toprule
	\multicolumn{2}{@{\hspace{1pt}}l}{
	  	\multirow{2}{*}{
	  		\begin{tabularx}{2.2cm}{@{\hspace{2.5pt}}l}
	  			\\[-11pt]Training \\[-2pt] augmentation
	  		\end{tabularx}
	  		}
		} 
		& \multicolumn{4}{c}{Test time augmentation}\\[1pt]
	 	\cmidrule(l){3-7}
		& & None & 1 & 5 & 20 & 100\\[1pt]
	\midrule
	\\[-12pt]
	None  & -- & 5.74 & 6.74 & 7.26 & 7.21 & 7.26 \\[1pt]
	Random & 5 & 39.58  & 39.79 & 54.05 & 59.84 &  62.68 \\[1pt]
	 & 20 & 54.05 & 52.84 & 63.21 & 67.68 & 68.47 \\[1pt]
	Orthogonal & 5 & 41.16 & 42.16 & 52.89 & 56.68 & 58.68
	\\[1pt]
	 & 20  & 56.16 & 54.74 & 66.47 & 71.00 & 71.84\cellcolor{LightCyan} \\
	\bottomrule
\end{tabularx}
\end{small}
\end{table}
We study the impact of multipath on fingerprinting using a Rayleigh fading model \cite{rappaport1996wireless} with $L$ multipath components:
\begin{equation*}
h(t)= \sum_{k=1}^{L} A_k e^{j\phi_k} \delta(t-\tau_k),
\end{equation*}
where
$A_k\sim\text{Rayleigh}\,(P_k)$, $\phi_k\sim\text{Uniform}\,(0, 2\pi)$ and $\delta(\cdot)$ is the Dirac delta function. We use the Extended Pedestrian A (EPA) profile, a well-known statistical channel model used in LTE system testing \cite{itu2012lte}. As shown in Table \ref{table_EPA}, this profile quantifies the delays $\tau_k$ and relative powers $P_k$ of the multipath components. 

\vspace{4pt}
\noindent \textbf{``Different day'' scenario (no augmentation or equalization):} We investigate training and testing on different emulated days similar to prior CFO experiments. 
Using the EPA profile, we use different realizations of the channel vector for each day and for each device.
Each realization has 7 multipath components chosen from a Rayleigh distribution with relative powers and delays specified in Table IV.  We do not vary the channel realization for a given device on a given day, hence we are modeling quasi-static environments.}
With single day training, we get excellent performance when testing on the same day (98\%), but very poor accuracy if we test on a different day (5.8\%). This clearly indicates a lack of robustness to channel variations, with the network involuntarily locking on to the channel as a means of discriminating between devices.

\vspace{5pt}
\noindent \textbf{"Different day" scenario with augmentation:} Assuming the received data is $f_\theta(\bxideal)$, we study the effect of channel augmentation ${\thetaaug}$ \textit{on top} of the emulated data: $f_{\thetaaug}(f_{\theta}(\bxideal))$.
We find that augmentation helps, but accuracy increases only to 47.8\% in the ``train on one day, test on another'' setting.
We can boost performance to 71.8\% if we are allowed access to training data over 2 emulated days (without increasing the size of the training set) and test on a third day, as shown in Table \ref{table_ch_dd}. 
Note that accuracy without augmentation is still low.
If training data spans 3 days, augmentation improves accuracy even further to 79.7\%.

This phenomenon can be understood by modeling channel variations in the frequency domain. Suppose transmitter $i$ sends message $X_i$ over ``physical'' channel $H_i$
\begin{equation*}
Y_i(f) = H_i(f)\, X_i(f),
\end{equation*}
and we augment with randomly chosen channels $G$:
\begin{align*}
\tilde{Y}_{i}(f) &= G(f)\, Y_i(f) \\&= G(f) \,H_i(f) \,X_i(f).
\end{align*}
The effective channel $G(f) \,H_i(f)$ will still contain all the nulls of $H_i$, which could potentially be correlated with the transmitter ID. Thus, augmentation alone cannot completely remove the effect of the underlying physical channel. Access to more varied training data, when combined with augmentation, increases the diversity of the overall channel that the network sees.


\begin{figure}[!t] 
  \centering
   \includegraphics[width=0.95\columnwidth]{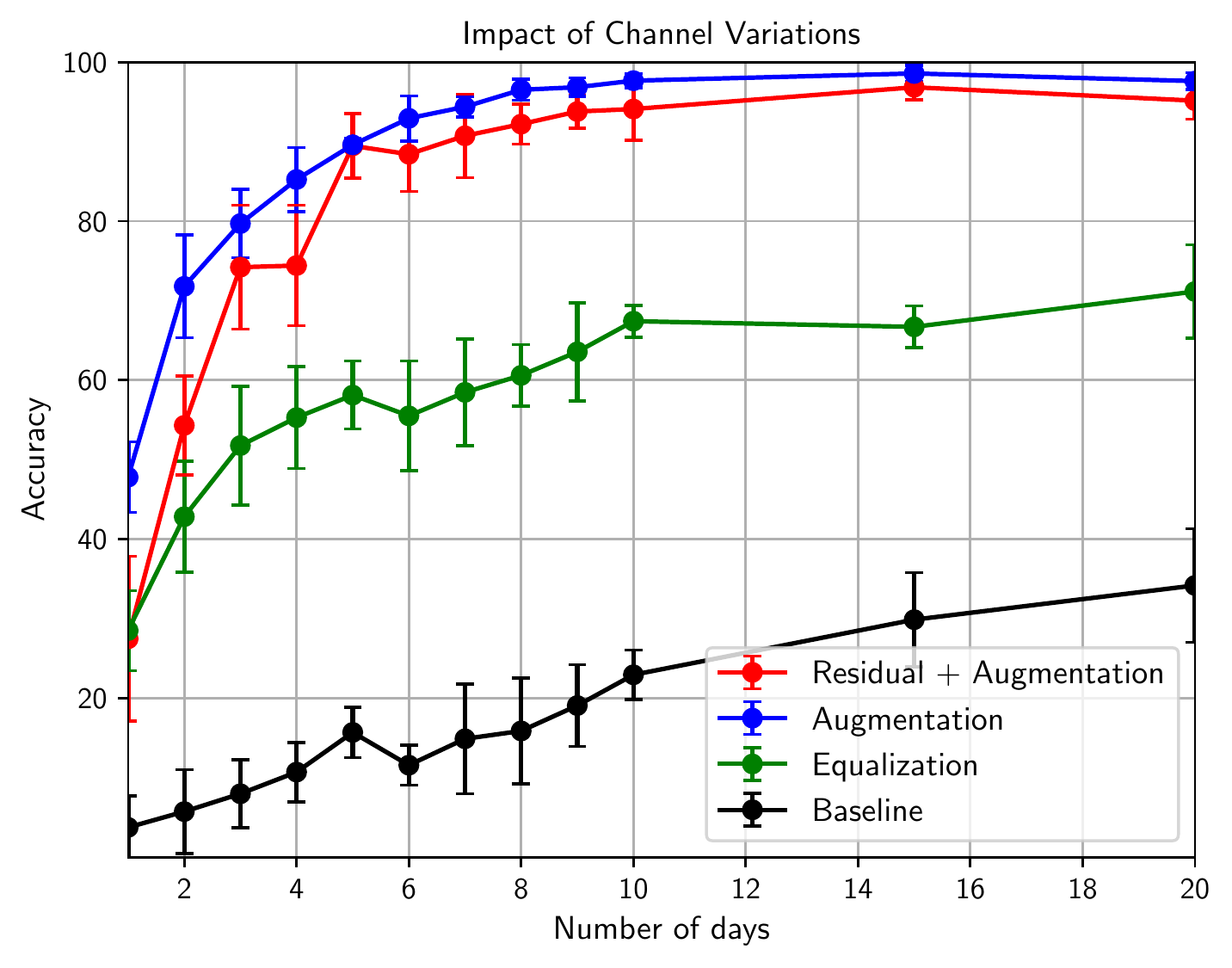}
  \caption{
  Comparison of channel equalization and augmentation as we increase the number of emulated days for training (with the size of the training set kept constant). 
  Baseline accuracies are reported for a network trained without augmentation or equalization.
  }
  \label{fig:chaug_eq}
  \vskip -1em
  
\end{figure}

The preceding results are achieved using 20 training and 100 test augmentations (with soft outputs added up over 100 augmented copies of each test packet). As before, we find that the ``orthogonal'' approach works the best for training: using the same set of channels across devices discourages the network from learning to use the channel as a fingerprint. 
Fig.\ \ref{fig:testaughist} illustrates the impact of test time augmentation on the distribution of soft outputs $p(\hat{y})$ for two sample devices. If we do not augment the test set, many samples from device 4 are misclassified as device 7 (shown in the first row of Fig.\ \ref{fig:0test}). As the number of test augmentations increases (Fig.\ \ref{fig:10test}, \ref{fig:100test}), we get increasingly precise estimates of the desired prediction \eqref{eq:testaug}, causing $p(\hat{y}=7|y=4)$ to shift towards $0$, and $p(\hat{y}=4|y=4)$ towards $1$.

\vspace{5pt}
\noindent \textbf{``Different day'' scenario with equalization:}
Another strategy to remove channel influence would be to equalize signals using the long training field of the WiFi preamble. We equalize data in the frequency domain and compare results with augmentation in Fig.\ \ref{fig:chaug_eq}.  Each experiment is performed with 5 different seeds, with error bars denoting one standard deviation from the mean.
We find that equalization performs much poorer than channel augmentation, with a performance gap of 26.5\% even with 20 training days. It appears that the residual distortion after equalization is large enough to swamp out the nonlinear characteristics that we are interested in. 

\begin{figure}[t]
  \centering
    \includegraphics[width=0.95\columnwidth]{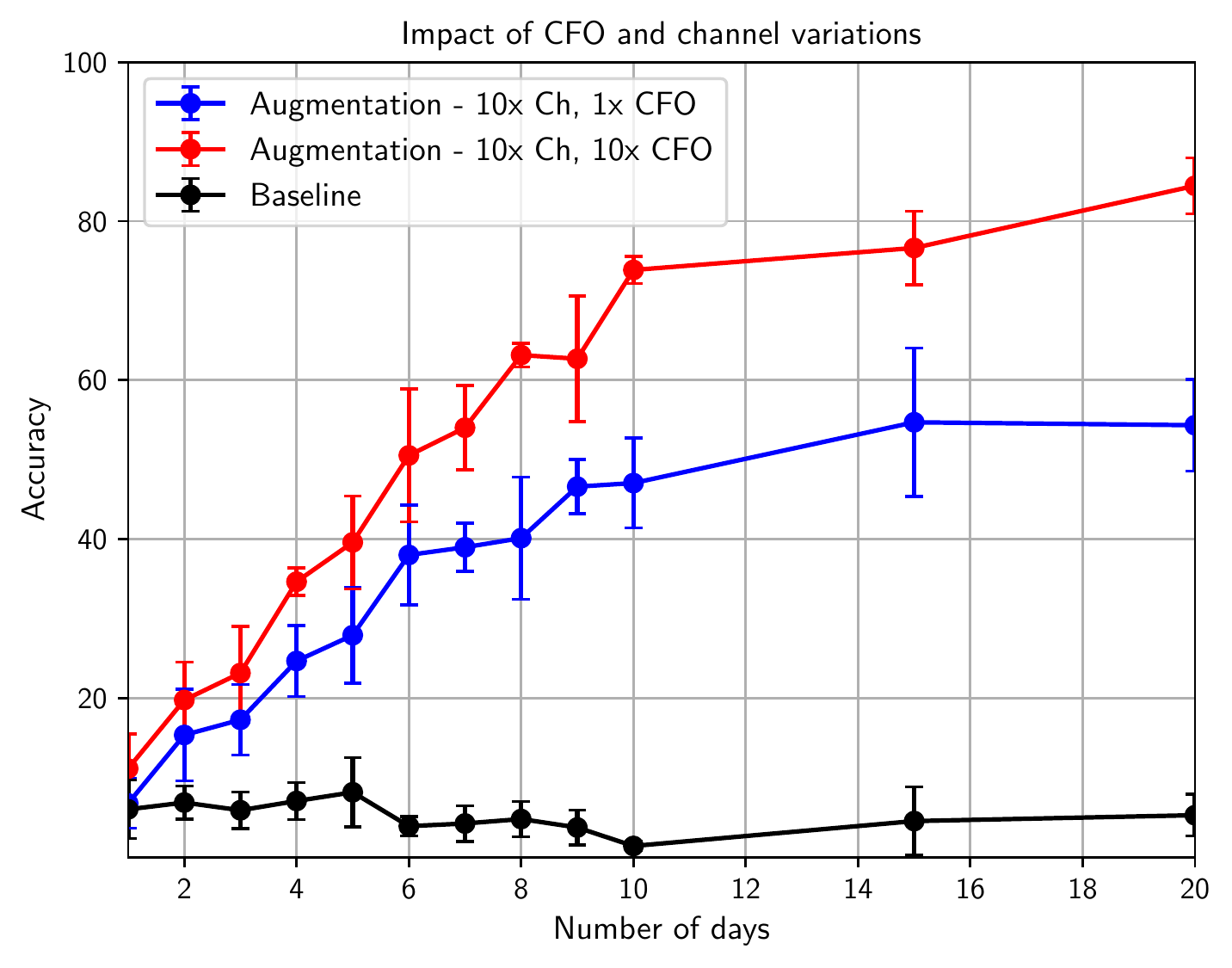}
  \caption{Performance of training augmentation across days when there is a combination of CFO and channel variations. We use the orthogonal augmentation approach for channels and the random method for CFOs.}
   \label{fig:ch_cfo}
  \vskip -.75em
\end{figure}

\begin{table}[!b]
\caption{Comparison of augmentation, estimation and the residual approach when both the CFO and channel vary.}
\label{table_cfo_ch_dd}
\vskip 0.05in
\centering
\begin{small}
\setlength\extrarowheight{1pt}
\begin{tabularx}{0.98\columnwidth}{lXXXX}%
	\toprule
	\multirow{2}{*}{\begin{tabular}{@{}l}\\[-6pt] Training strategy \end{tabular}}
	 & \multicolumn{4}{c}{Number of days}\\[1pt]
	\cmidrule(l){2-5}
	& 2 & 5 & 10 & 20\\[1pt]
	\midrule
	\\[-12pt]
	Residual + augmentation & 19.11 & 26.21 & 67.50 & 78.95  \\[1pt]
	Pure augmentation 		   & 24.90 & 49.36 & 77.83 & 90.10  \\[1pt]
	CFO comp. + channel aug.    & \cellcolor{LightCyan}33.96 & \cellcolor{LightCyan}62.63 &\cellcolor{LightCyan} 88.96 & \cellcolor{LightCyan}91.40 \\
	\bottomrule
\end{tabularx}
\end{small}
\end{table}

\vspace{5pt}
\noindent \textbf{Residual approach:} As previously described (Section \ref{sec:augmentation}), we can use the estimated channel to obtain residual noise and use it as CNN input. When combined with augmentation, we obtain accuracies that are competitive with, but not better than, pure augmentation, as shown in Fig.\ \ref{fig:chaug_eq}. 
We speculate that errors in channel estimation prevent the residual method from offering a clear advantage in accuracy, in contrast to the simpler setting of CFO uncertainty considered in Section \ref{sec:cfo}.

Overall, augmentation is the best of the three considered strategies for making networks insensitive to channel effects: with 10 training days, it can restore accuracy to 97.7\%.



\begin{figure}[!t]
  \centering
  \begin{subfigure}[t]{.9\columnwidth}
    \includegraphics[width=\textwidth]{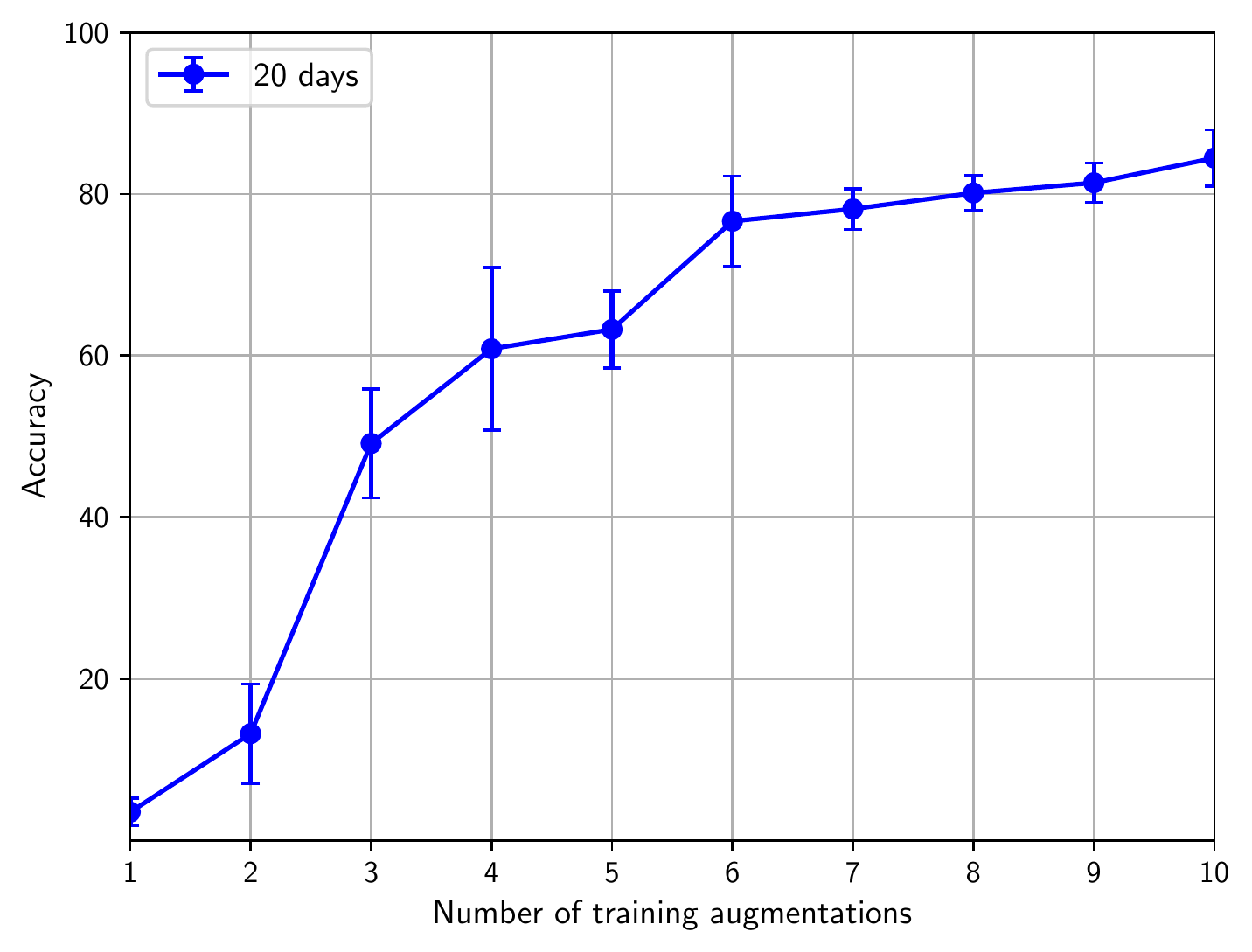}
    \caption{Effect of increasing training augmentations.}
    \label{fig:ch_cfo_numaug}
    \vspace{10pt}
  \end{subfigure}
  \begin{subfigure}[b]{.9\columnwidth}
    \includegraphics[width=\textwidth]{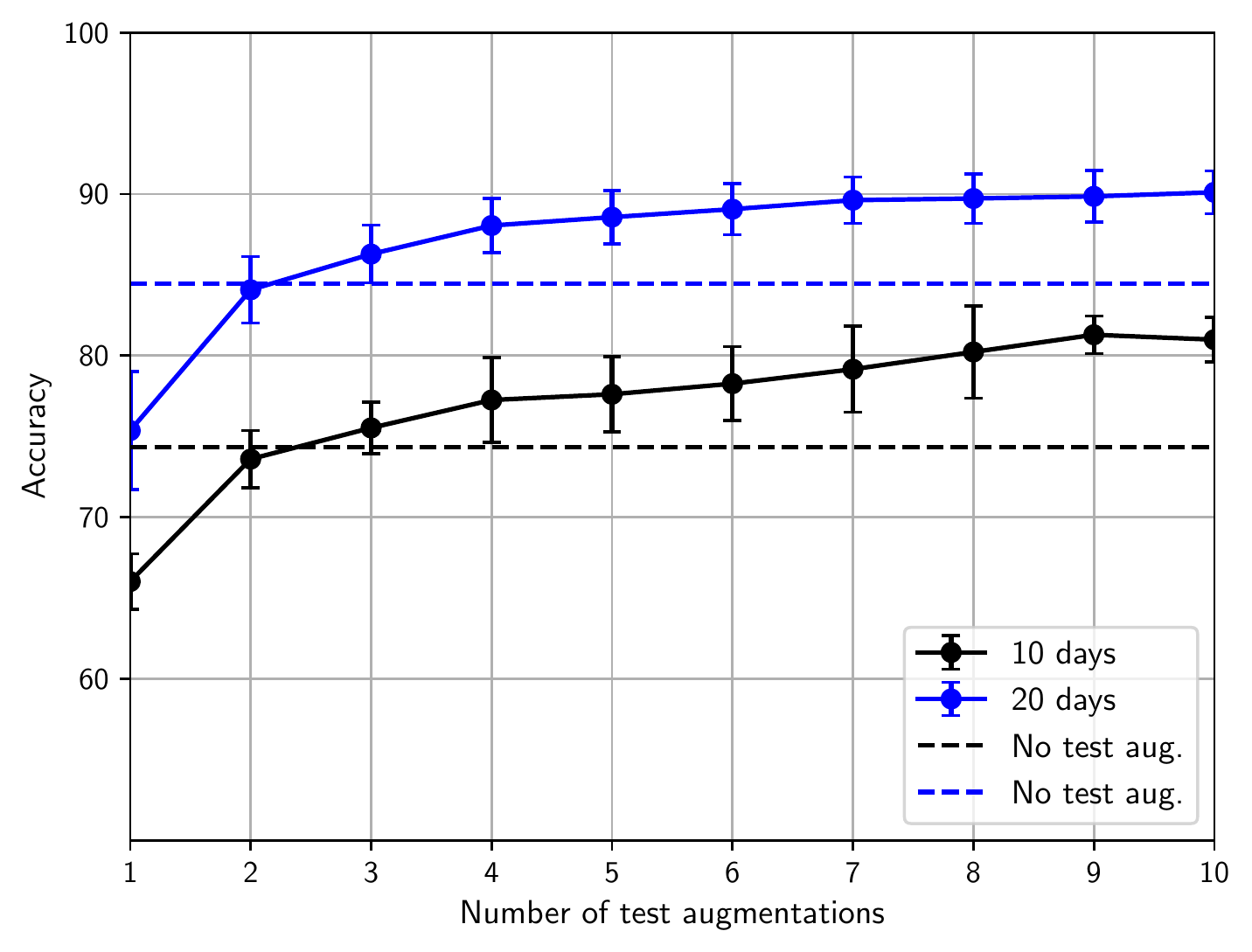}
    \caption{Effect of increasing test augmentations.}
    \label{fig:ch_cfo_testaug}
    \vspace{5pt}
  \end{subfigure}
  \caption{
  Accuracy as a function of the amount of augmentation when both the CFO and channel fluctuate. We augment the CFO and channel by equal amounts, with the $x$-axis denoting the number of augmentations for each.}
  \label{fig:ch_cfo_numaug}
  \vskip -.75em
\end{figure}

\begin{figure*}[t]
\centering
\includegraphics[width=0.98\textwidth]{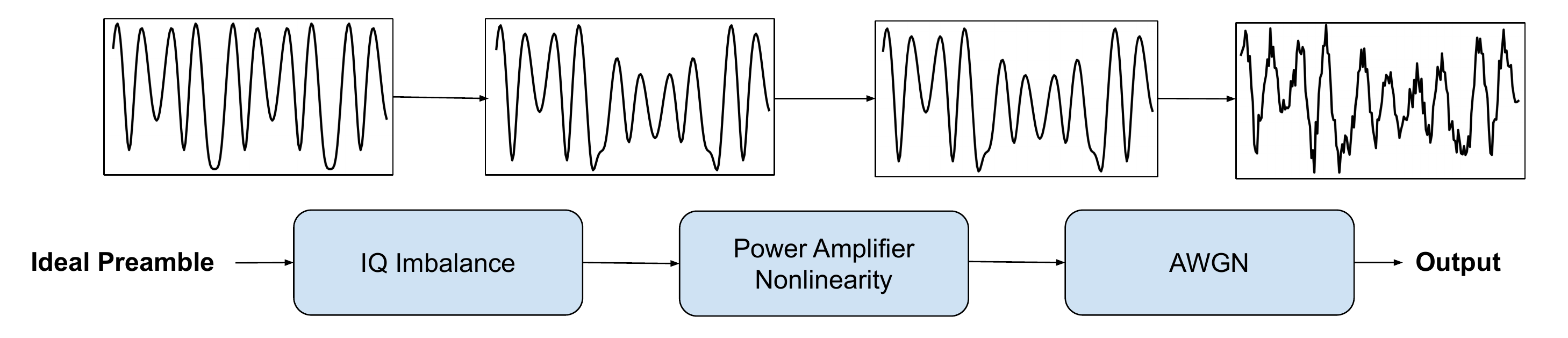}
\caption{Block diagram for generation of the simulation-based dataset.}
\label{fig:nonlinearity_sim}
\vskip -0.1in
\end{figure*}

\subsection{Combination of Channel and Carrier Offsets} \label{sec:cfo_channel}

Lastly, we focus on a combination of channel and carrier offsets across different days. This is a harsher and more realistic setting than prior experiments, with test set accuracy without augmentation or compensation no better than random guessing (5\%) even if training data spans 20 emulated days. 

\vspace{5pt}
\noindent\textbf{Augmentation:}
We explore data augmentation with randomly generated channels and CFOs, and report results in Figures \ref{fig:ch_cfo} and \ref{fig:ch_cfo_numaug}. We find an equal number of augmented CFOs and channels to work well: when using 20 training days, performance improves from 5\% to 84.4\% with training augmentation alone, and to 90.1\% with both training and test augmentation. We observe that the amount of test augmentation is important: as shown in Fig. \ref{fig:ch_cfo_testaug}, if we only augment test data 2 times, we observe a drop in accuracy. This is because the Bayesian average \eqref{eq:testaug} requires a large number of realizations of the two nuisance parameters (CFO, channel) in order to be accurate.

\vspace{5pt}
\noindent\textbf{Estimation:}
Table \ref{table_cfo_ch_dd} reports on comparisons with estimation strategies, the residual approach and also a mix of estimation and augmentation. We find that equalization, when combined with either CFO compensation or augmentation, results in only 10\% accuracy and therefore do not include it in the comparison. The best result is obtained by a combination of CFO compensation and channel augmentation for both training and test sets, with competitive performance from pure augmentation when the number of days of training is large.


\subsection{Simulation-Based Dataset} \label{sec:simulated_dataset}

Since the datasets used in the previous sections are not publicly available, in the interest of reproducibility and as a contribution to the community, we have created a simulation-based WiFi dataset \cite{repo} based on models of some typical nonlinearities \cite{schenk2008rf,razavi2012rf,wifi1999}. We implement two different kinds of circuit-level impairments: I/Q imbalance and power amplifier nonlinearity, with Figure \ref{fig:nonlinearity_sim} depicting the order in which the nonlinear effects were added. We skip effects of the digital to analog converter such as DNL and INL. In a manner similar to prior sections, we perform experiments to study the effect of channel and CFO variations on fingerprinting performance. We now discuss the models and parameters used to generate the nonlinear effects.

\begin{table}[b]
\caption{Fingerprinting performance on the simulated dataset in the ``different day'' scenario for both CFOs and channels, when using 20 days for training.}
\label{table_simulated}
\vskip 0.05in
\centering
\centering
\begin{small}
\begin{tabular}{lcccccc}%
  \toprule
  \multirow{2}{*}{{Training strategy}}
  & \multicolumn{3}{c}{Test time augmentation} \\
  \cmidrule(l){2-4} 
  & None & 1 & 100 \\[1pt]
  \midrule
  No augmentation or compensation & 7.61 & 6.68 & 8.30  \\[1pt]
  Pure augmentation        & 81.38 & 77.56 & 86.24  \\[1pt]
  CFO comp. + channel aug.    & 81.59 & 81.98 & 91.80 \\
  \bottomrule
\end{tabular}
\end{small}
\end{table}

\vspace{0.2cm}
\noindent \textbf{I/Q Imbalance:} The I/Q imbalance \cite{schenk2008rf} can be modeled as follows, with parameters $\epsilon$ and $\phi$ representing gain and phase mismatch respectively: 
\begin{multline*}
\tilde{s}_{\mathrm{RF}}(t) = \sc(t) \left(1+\frac{\epsilon}{2}\right) \cos(2\pi \fc t +\frac{\phi}{2} ) \\
- \ss(t) \left(1-\frac{\epsilon}{2}\right) \sin(2\pi \fc t -\frac{\phi}{2}).
\end{multline*}
Since the IEEE 802.11 WiFi standard \cite{wifi1999} specifies an error vector magnitude (EVM) of $-19$ dB, we set $\epsilon\leq0.2$ and $|\phi|\leq\pi/30$. In order to simulate 19 different devices (similar to original dataset) we choose distinct $\epsilon$ values for each device from the set $[0,0.2]$ uniformly, i.e.\ \{$0, 0.2/19, 0.4/19 ...$\}. Similarly, we pick $\phi$ from the set $[-\pi/30,\pi/30]$ uniformly. We note that all the values are shuffled randomly before matching to each device, hence extreme cases for both parameters are most likely not on the same device.

\vspace{0.2cm}
\noindent \textbf{Power Amplifier Nonlinearity:} The power amplifier (PA) is another source of circuit-level nonlinearity that varies across devices. There are a number of different models for this nonlinearity \cite{Saleh81, Zhu2004Volterra, Ku2003behavioral, Pedro2005overview}. We model PA nonlinearities as a saturated third-order polynomial function \cite{razavi2012rf}: 
\begin{equation*}
y(t) = \begin{cases} x(t)\cdot\left(1- \dfrac{0.44|x(t)|^2}{3\pdb}\right) &\mbox{if } |x(t)|^2 \leq \dfrac{\pdb}{0.44}, \\ 
\dfrac{x(t)}{|x(t)|}\sqrt{\pdb} & \mbox{if } |x(t)|^2 > \dfrac{\pdb}{0.44}. \end{cases}
\end{equation*}
This function is parametrized by the 1 dB compression point $\pdb$, defined as the output power level at which the gain decreases 1 dB from its constant value.
Similar to I/Q imbalance, we determine the range of the values for $\pdb$ that satisfy the EVM specifications. We choose $\pdb$ values for each device uniformly from the set $[8.45,20]$. 
The corresponding transfer functions are depicted in the appendix.

\vspace{0.2cm}
\noindent \textbf{Adding AWGN:}
After obtaining preamble signals with nonlinear features for 19 different devices, we create training, validation and test datasets by adding additive white Gaussian noise (AWGN) such that $\text{SNR}=20$ dB for each dataset. For training, we use 200 signals per device from 19 devices. The validation and test sets contain 100 signals per device. Overall, the dataset contains 3800 signals for training, 1900 signals for validation and 1900 signals for the test set.

\vspace{0.2cm}
\noindent \textbf{Results:}
We use the same CNN and training hyperparemeters as before, except for the number of epochs, which we set to 100. We observe trends similar to our results on emulation of ``different days'' with the measured WiFi data: model-based augmentation can significantly help improve performance when training over multiple emulated days and testing on a different day. We report on these results in Table \ref{table_simulated}.


\section{Conclusions}

While complex-valued CNNs are a promising tool for learning RF signatures, we conclude that blind adoption of these networks is dangerous due to confounding factors that impede generalization across space and time. We show that model-based augmentation is a useful tool for handling such confounding factors; a novel finding is that 
augmentation is helpful not just for training, but also during inference. A lower-complexity alternative to augmentation is to estimate and undo the effects of confounding 
factors using detailed, protocol-specific models, but, depending on the phenomenon of interest, the residual errors (e.g., from channel estimation) 
may swamp out the weaker nonlinear effects that we wish to learn.  A judicious combination of estimation and augmentation can confer robustness,
but augmentation alone is a competitive approach when we seek protocol-agnostic strategies.

Our results highlight the promise and pitfalls of deep learning for RF signatures, rather than providing definitive answers.
There are a number of open issues for further investigation, including alternative DNN architectures and fundamental detection-theoretic limits to provide benchmarks for robust fingerprinting,
Another important area for future work is exploration of the robustness of DNN-based RF signatures to adversarial attacks.  Adversarial attacks and defenses are a topic of intensive investigation in the context of standard image datasets \cite{goodfellow2014explaining,madry2017towards,athalye2018obfuscated}, but it is of interest to explore threat models that are specifically tailored to wireless physical layer security.  Finally, it is important to investigate RF and mixed signal circuit design issues associated with the concept of RF signatures, including the potential for deliberately introducing manufacturing variations to enable discrimination, and characterization of the stability of device nonlinearities to environmental variations (e.g., in temperature and moisture).


\section*{Acknowledgment}

This work was funded in part by DARPA under the AFRL contract number FA8750-18-C-0149, by ARO under grant W911NF-19-1-0053, and by the National Science Foundation under grants CNS-1518812 and CIF-1909320. The views and conclusions contained herein are those of the authors and should not be interpreted as necessarily representing the official policies or endorsements, either expressed or implied, of DARPA or Air Force Research Laboratory or ARO or the U.S. Government. The authors gratefully acknowledge research discussions with collaborators at Teledyne Scientific, including Mark Peot, Laura Bradway, Karen Zachary and Michael Papazoglou.

{
\footnotesize{
\bibliographystyle{IEEEtranN}
\bibliography{references}
}}

\appendices


\begin{appendices}

\section{Training details}

Networks are trained for 200 epochs with a batch size of 100, using the Adam optimizer with learning rate $\eta=0.001$ and weight decay constant $\lambda=0.0001$. We normalize all signals to unit power. 
  For weight initialization, we use the complex-valued Glorot initialization from \cite{trabelsi2017deep} for complex layers, and the real-valued Glorot \cite{glorot2010understanding} for real layers. Detailed information about network architecture can be found in Tables \ref{table:arch_adsb_complex} and \ref{table:arch_wifi_complex}.
For all experiments, we use Keras \cite{chollet2015keras} with Theano backend, since complex-valued layers are implemented in Keras. We use the NVIDIA GeForce GTX 1080Ti GPU and observe that an epoch of training takes about 0.8 seconds, when using the WiFi data with 200 samples per device (from 19 devices). 

To assess performance, we have used the average of 5 different runs with different random seeds for initial weights and with different random realizations of CFOs and channels used for emulation and augmentation. In all the graphs in Section V, error bars denote one standard deviation from the mean over different runs. Confusion matrices are reported in Fig.\ \ref{fig:confusion}. Table \ref{table_simulated2} provides more details on performance for the simulated dataset, reporting the means and standard deviations for all scenarios.
We have also carried out 5-fold cross validation, where we use 5 different randomly chosen partitions of the data for training and testing, with the result that there is very little variation in performance. We provide an example result: when we use stratified 5-fold cross validation for the 20 day channel experiment, using data augmentation only on training set, we obtain test accuracies of 91.42\%, 91.58\%, 85.95\% 91.47\%, 96.58\%.
(Since there is no test time augmentation for this particular result, we note that these numbers are slightly lower than the numbers reported in Figure \ref{fig:chaug_eq}).

\section{Comparison of complex and real networks}

We compare the performance of complex-valued and real-valued networks in Table \ref{table_complex_reim}. For real networks, we follow the approach of \cite{o2016convolutional,o2017introduction,sankhe2018oracle} in treating real and imaginary parts of input data as different channels. For a fair comparison, we consider real networks with different scaling factors for the number of channels (the numbers in brackets in Table \ref{table_complex_reim}). This is to account for the fact that a complex filter would contain twice as many parameters as an equivalent real filter. Since the last two layers of the complex network are real-valued, we do not scale the corresponding layers of the real network. We find that the complex network outperforms all its real counterparts, with a performance gain of 6.6\% for ADS-B and 1.6\% for WiFi.

\begin{table}[b]
\centering
\caption{Performance comparison between complex-valued and real-valued networks. The scaling factor in brackets refers to the scaling for the number of channels.}
\label{table_complex_reim}
\begin{center}
\begin{small}
\begin{tabular}{llcr}
\toprule
Dataset & Network type & Accuracy & 
\begin{tabular}{@{}r@{}} Total number of \\ real parameters \end{tabular} 
\\
\midrule
ADS-B & Complex & \textbf{81.66} & 128,400$\phantom{iia}$\\
& Real & 73.84 & 78,400$\phantom{iia}$\\
& Real (1.4x) & 73.25 & 133,680$\phantom{iia}$\\
& Real (2x) & 75.00 & 246,600$\phantom{iia}$\\
\midrule
WiFi & Complex & \textbf{99.62} & 262,719$\phantom{iia}$\\
& Real & 97.50 & 162,319$\phantom{iia}$\\
& Real (1.4x) & 97.61 & 278,399$\phantom{iia}$\\
& Real (2x) & 97.94 & 512,519$\phantom{iia}$\\
\bottomrule
\end{tabular}
\end{small}
\end{center}
\end{table}

Architecture details for the complex and real CNNs we use are reported in Tables \ref{table:arch_adsb} and \ref{table:arch_wifi}, specifying the size and number of parameters in each layer for all the networks considered. Kernel sizes are specified using the notation [convolution size, number of input channels, number of output channels]. For real networks, the scaling factor in brackets refers to the scaling for the number of channels. Since the last two layers of the complex network are real-valued, we do not scale the corresponding layers of the real network. 
In order to prevent overfitting, in real-valued networks we use dropout \cite{srivastava2014dropout} with drop probability $p=0.5$ after fully connected layers, and weight decay with $\ell_2$ norm regularization parameter $\lambda=0.0001$.

\section{Visualizations}

Figure \ref{fig:visualize_layer_2} depicts input signals that strongly activate filters in the first and second layer of the ADS-B architecture. Since device-specific nonlinear effects manifest primarily as short-term transitions of amplitude and phase, the filters in the first layer can capture these effects by spanning a small multiple of the symbol interval (2 symbols). 
To compute these signals, we start from randomly generated noise and use 200 steps of gradient ascent to maximize the absolute value of each filter output, with the signal normalized to unit power at each step.

Transfer functions for the simulated power amplifier nonlinearities in Section IV-E are shown in Figure \ref{fig:pa_sim}.
The clean WiFi dataset was collected in a controlled indoor setting over the air. The data was analyzed via demodulation and channel estimation (using the preamble), with the observation that the channel is mostly flat, as shown in Fig. \ref{fig:ch_freq}. 

\begin{figure}[h]
\centering
\includegraphics[width=0.40\textwidth]{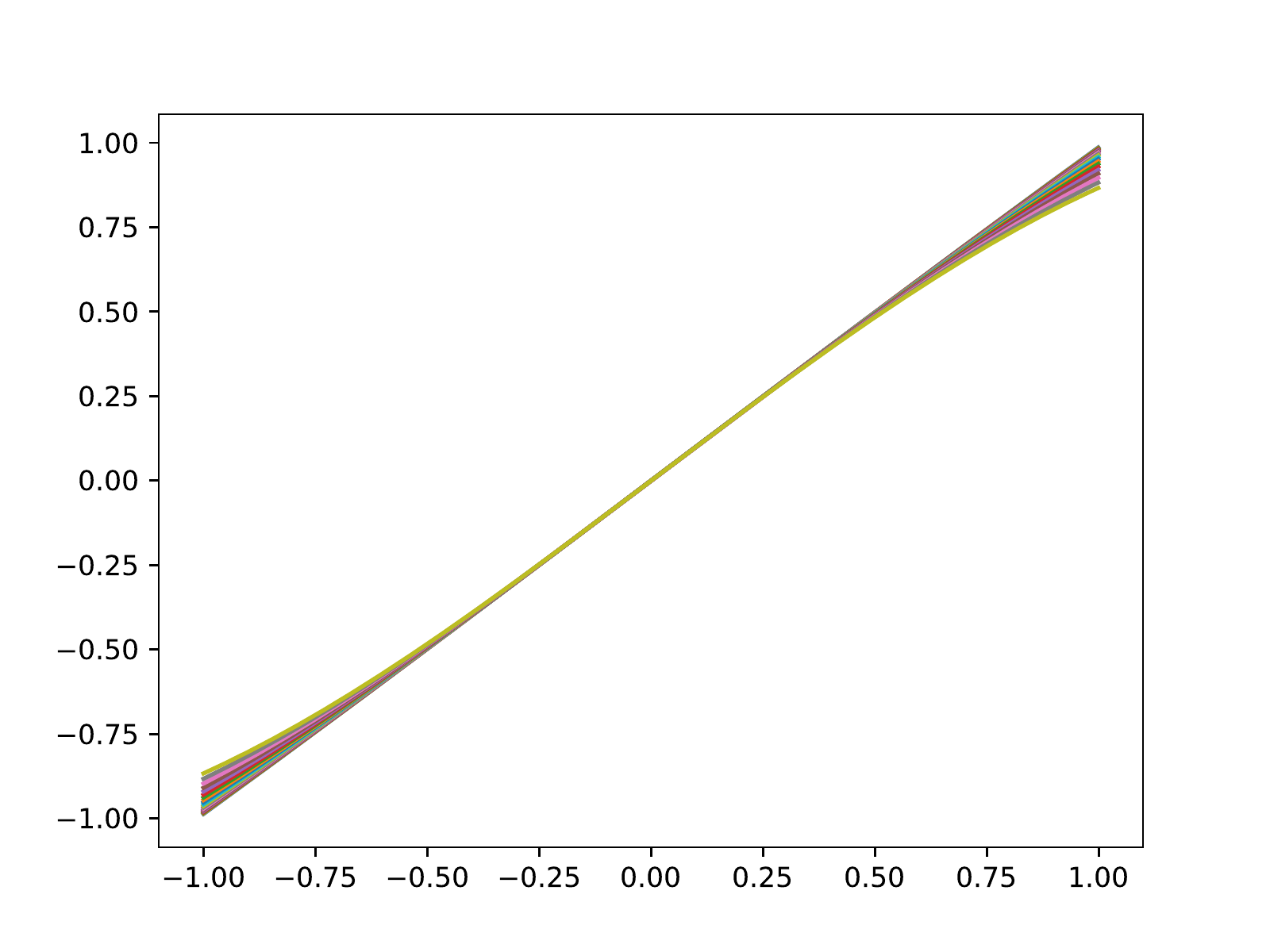}
\caption{Simulated power amplifier nonlinearities for different devices.}
\label{fig:pa_sim}
\vskip -0.1in
\end{figure}

\begin{figure}[h]
\centering
\includegraphics[width=0.40\textwidth]{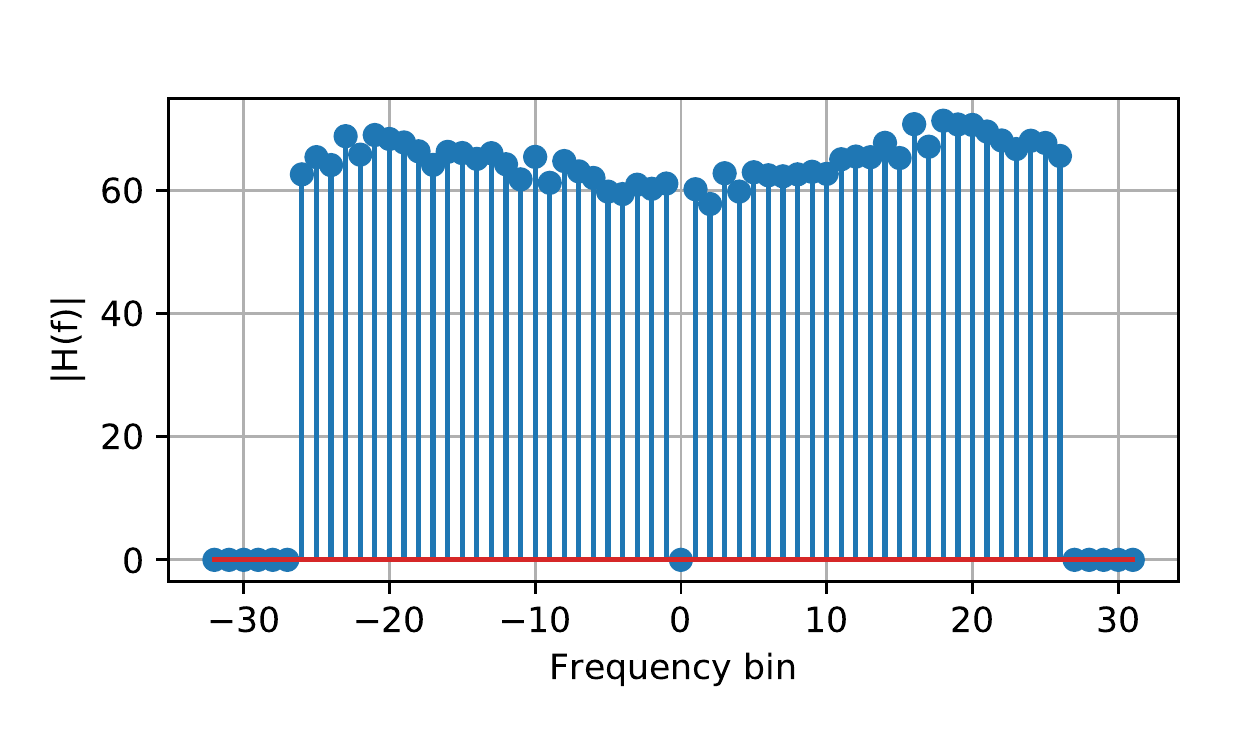}
\caption{Estimated channel frequency response of a sample signal from the clean WiFi dataset.}
\label{fig:ch_freq}
\vskip -0.1in
\end{figure}


\begin{figure*}[t]
  \centering
  \begin{subfigure}[b]{0.28\textwidth}
    \centering
  \includegraphics[width=1.03\textwidth]{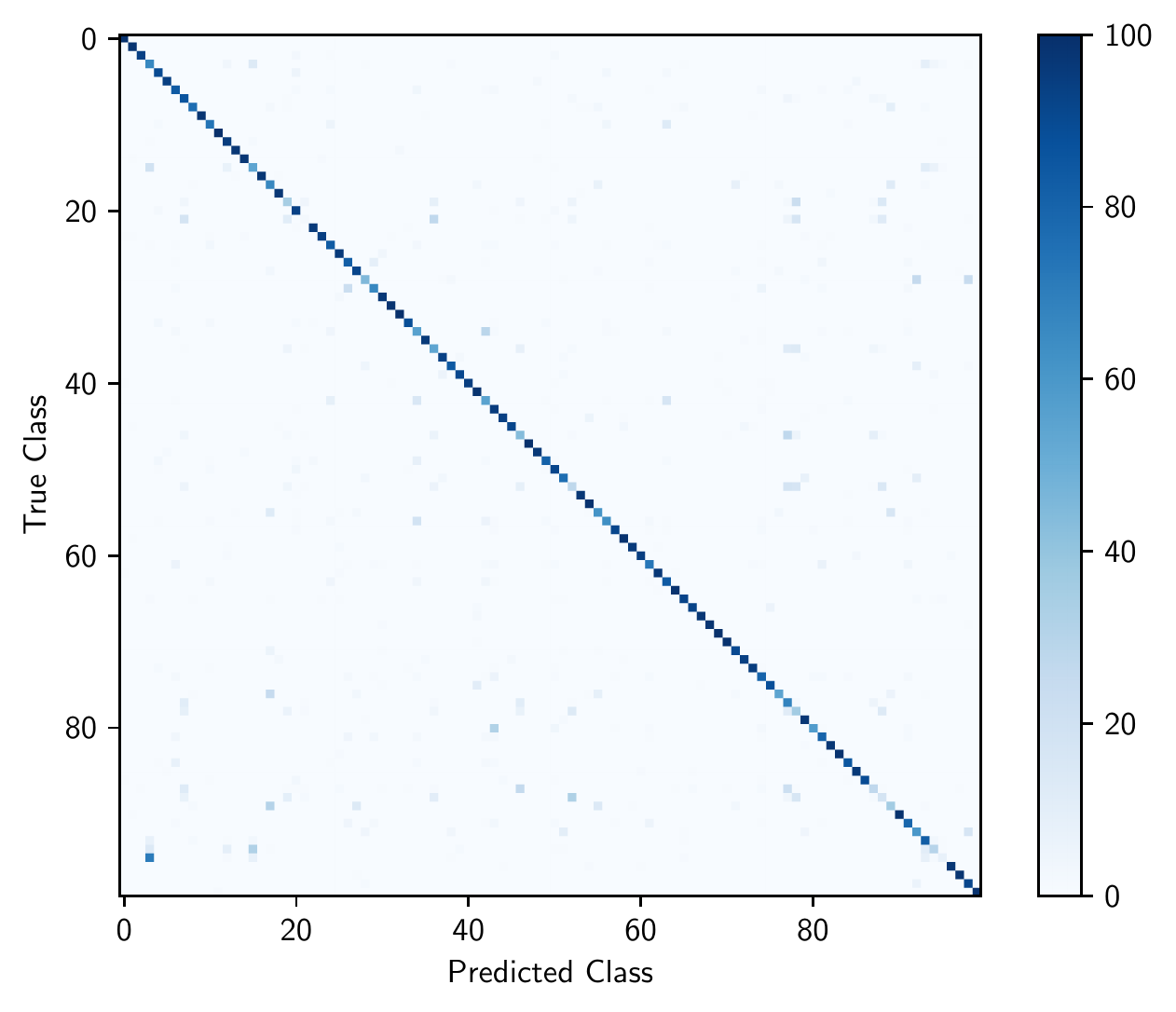}
  \caption{}
  \label{fig:conf_adsb}
  \end{subfigure}
  \hfill
  \begin{subfigure}[b]{0.28\textwidth}
    \centering
  \includegraphics[width=1.3\textwidth]{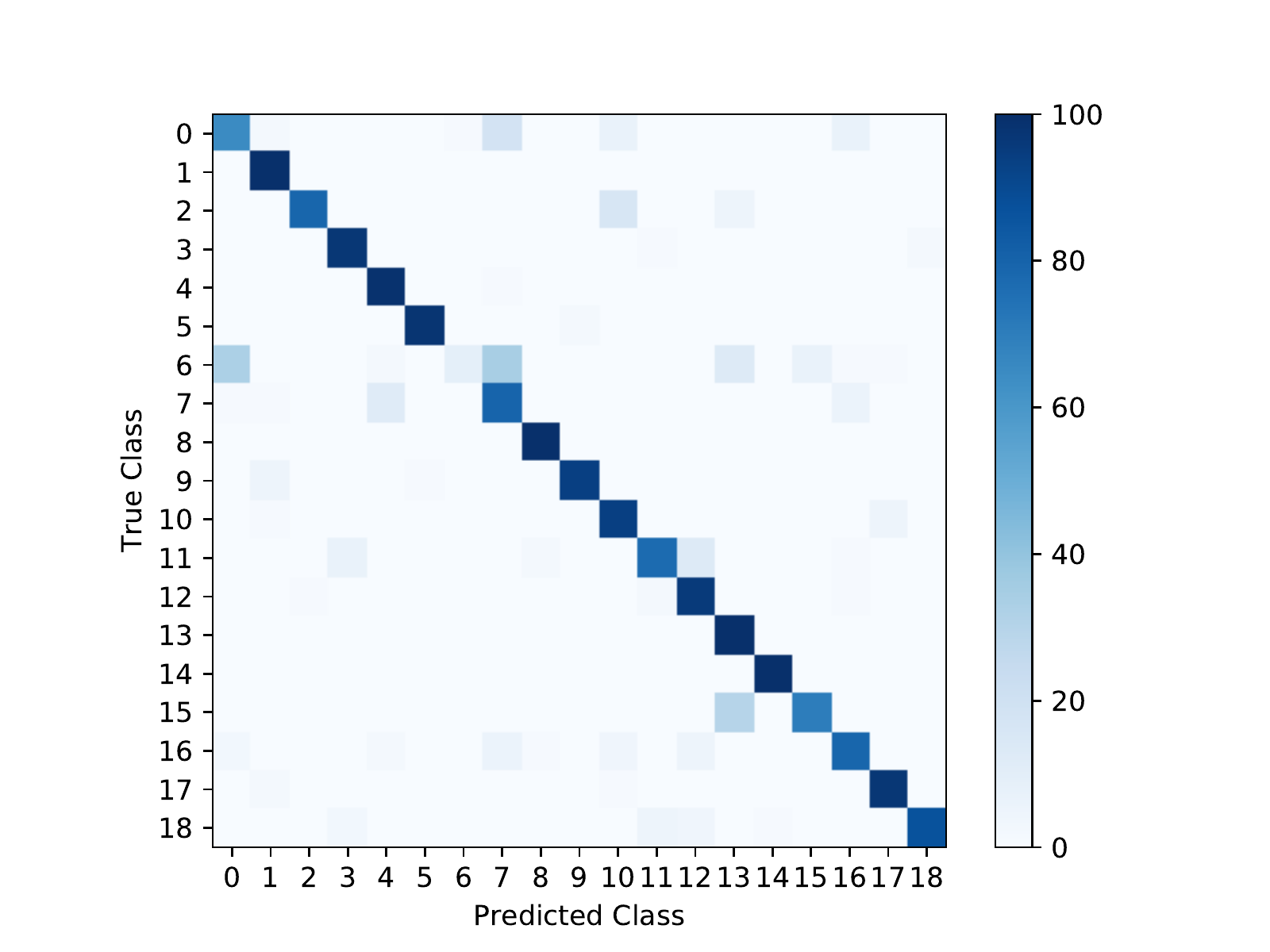}
  \caption{}
  \label{fig:conf_ch}
  \end{subfigure}
  \hfill
  \begin{subfigure}[b]{0.28\textwidth}
    \centering
  \includegraphics[width=1.3\textwidth]{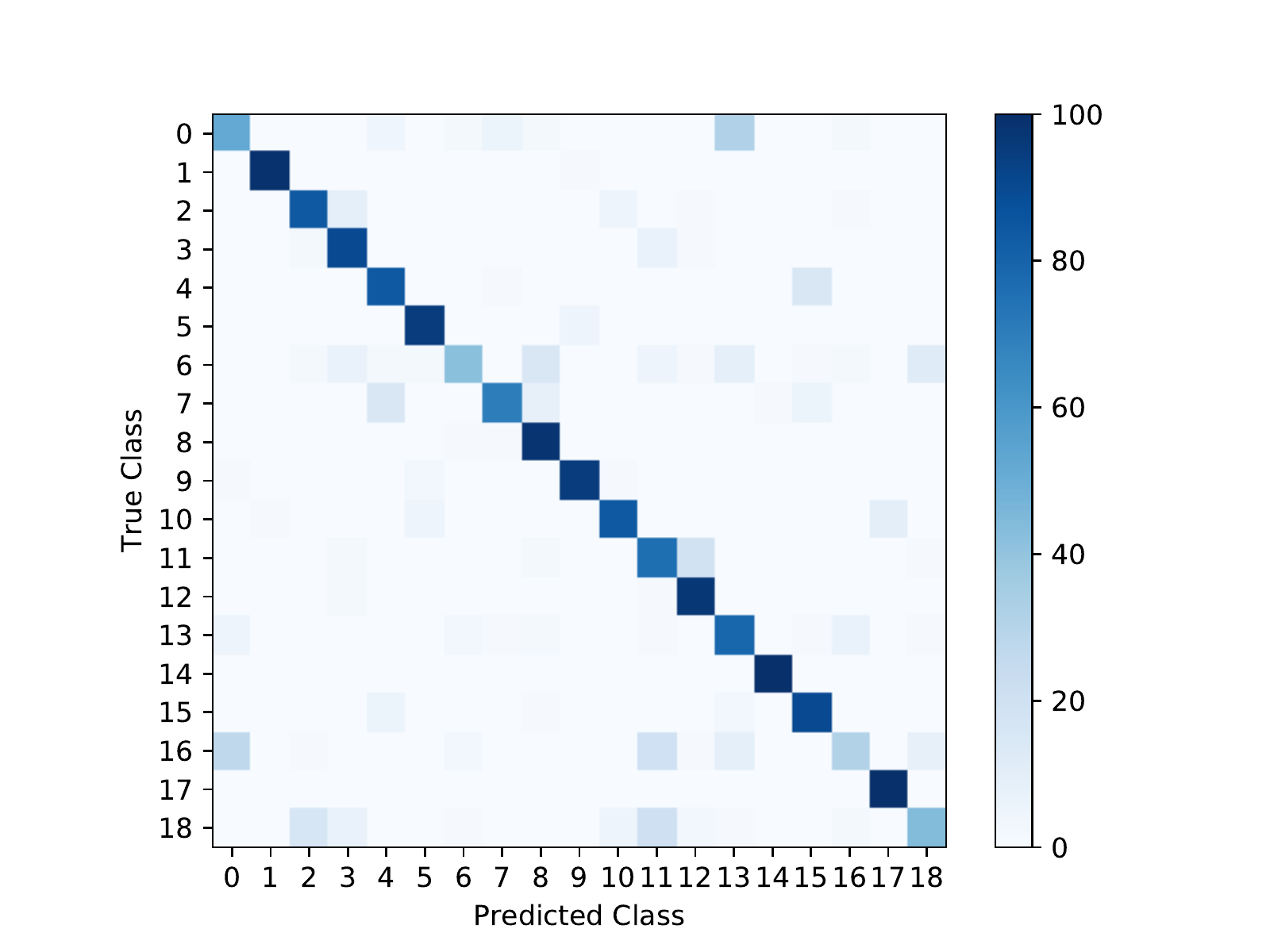}
  \caption{}
  \label{fig:conf_ch_cfo}
  \end{subfigure}
  \hfill
  \caption{Confusion matrices for fingerprinting of (a) the ADSB dataset (100 devices), (b) the clean WiFi dataset in the ``different day'' channel scenario (19 devices)  (c) the clean WiFi dataset in the ``different day'' channel + CFO scenario  (19 devices). For both (b) and (c), we use 20 days for training and a different day for testing, and perform 10 training augmentations.}
  \label{fig:confusion}
\end{figure*}

\begin{figure*}
\centering
\hspace{3pt}
\begin{subfigure}{0.265\linewidth}
    \includegraphics[width=0.48\linewidth]{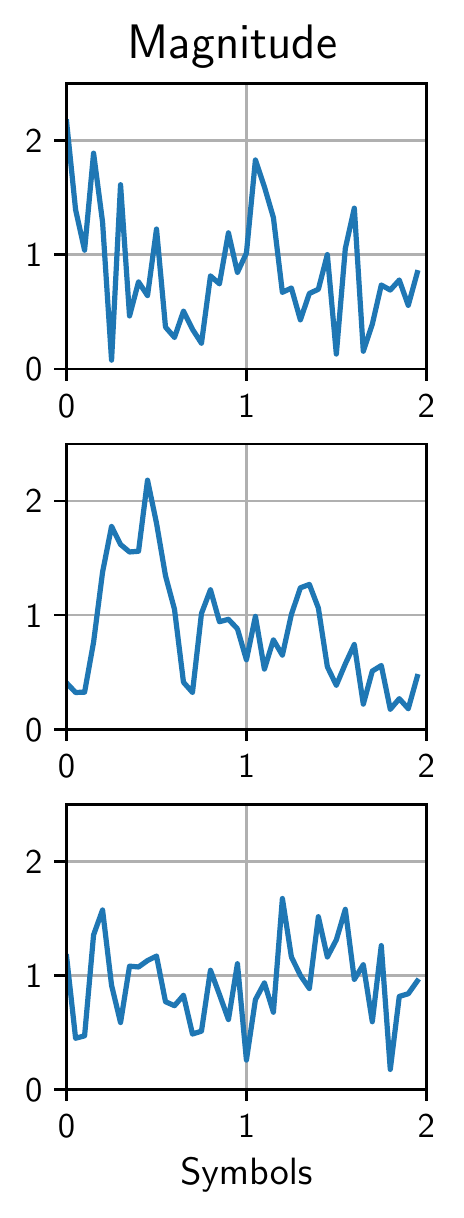}
    \includegraphics[width=0.48\linewidth]{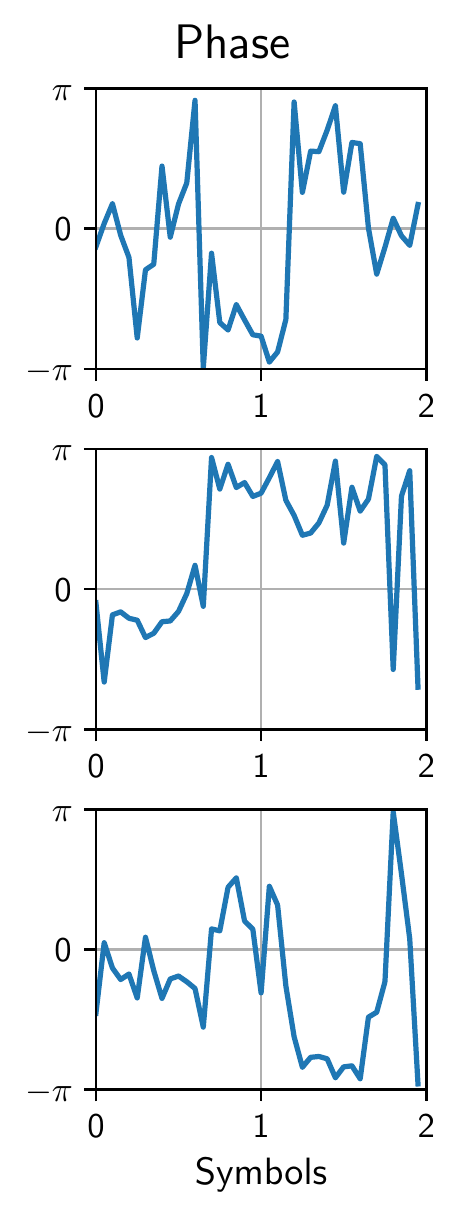}
    \caption{Layer 1}
    \label{fig:layer1} 
\end{subfigure}
\hfill
\begin{subfigure}{0.68\linewidth}
    \includegraphics[width=0.48\linewidth]{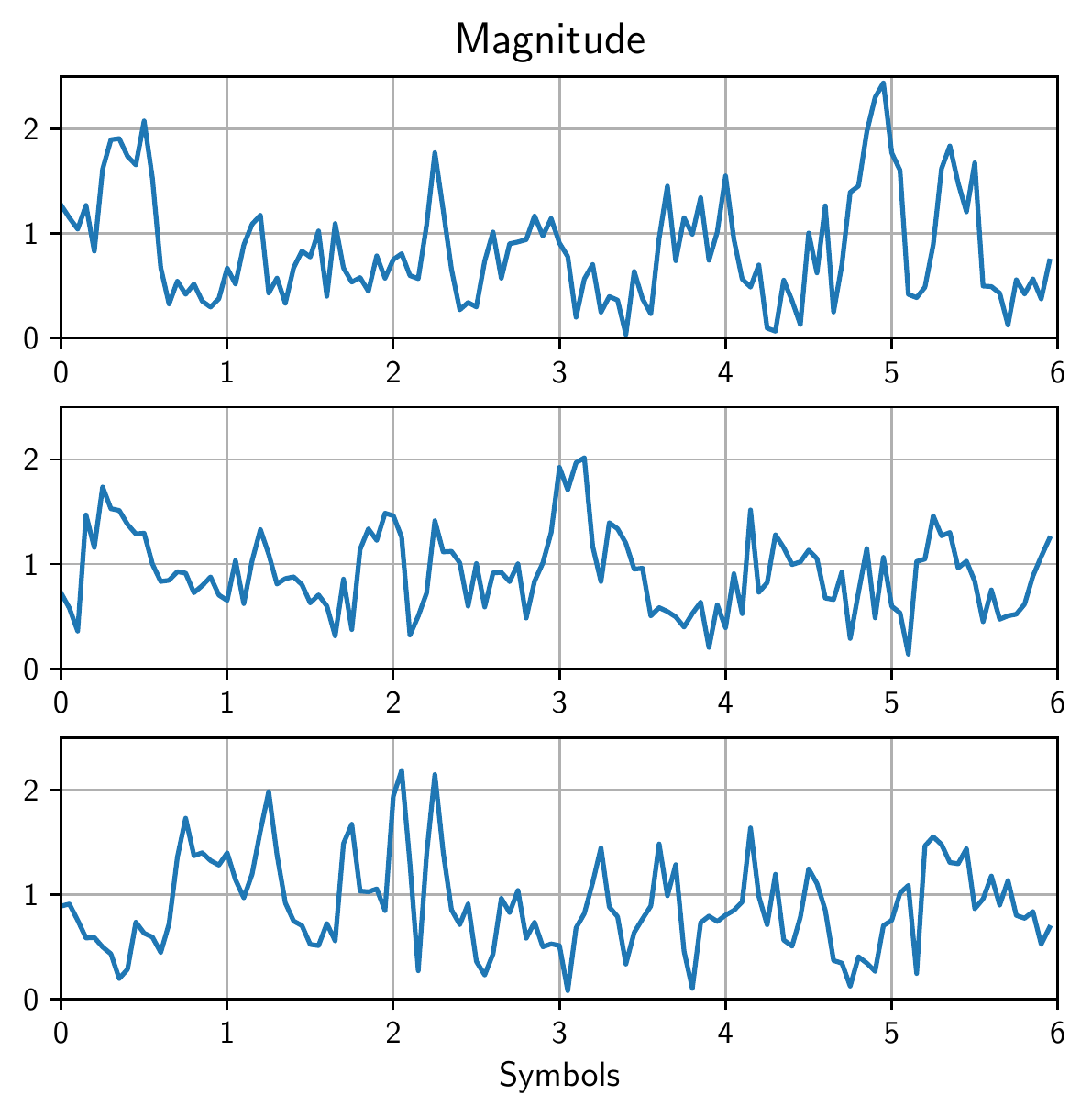}
    \includegraphics[width=0.48\linewidth]{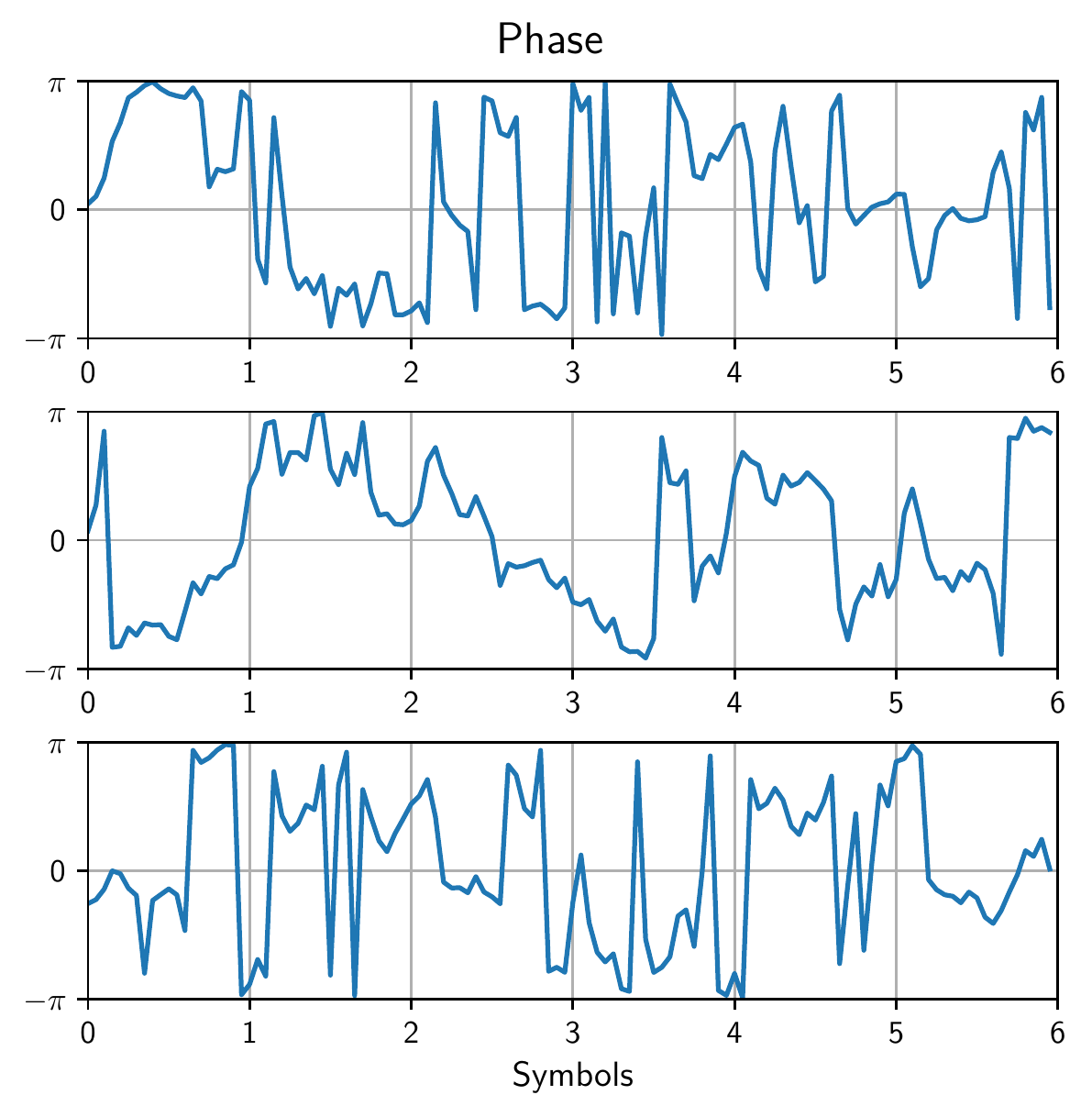}
    \caption{Layer 2}
    \label{fig:layer1} 
\end{subfigure}
\hspace{2pt}
\caption{Visualizations of the first and second convolutional layer for ADS-B (ModReLU architecture).
Each row shows the input signal that maximizes the activation of a particular filter, computed using gradient ascent starting from random noise (with signals normalized to unit power at each step). 
Convolutional filters in the first layer span 2 input symbols; filters in the second layer span 6 symbols.
}
\label{fig:visualize_layer_2}
\end{figure*}

\begin{table*}
\caption{Fingerprinting performance on the simulated dataset in the ``different day '' scenario for both CFOs and channels.}
\label{table_simulated2}
\vskip 0.05in
\centering
\begin{subtable}{\linewidth}
\caption{Performance when we use 20 days for training, and then test on a different day.}
\centering
\begin{small}
\begin{tabular}{lcccccc}%
  \toprule
  \multirow{2}{*}{{Training Strategy}}
  & \multicolumn{3}{c}{Test time Augmentation} \\
  \cmidrule(l){2-4} 
  & None & 1 & 100 \\[1pt]
  \midrule
  No aug. or comp. & 7.61$\pm$3.83 & 6.68$\pm$1.76 & 8.30$\pm$4.78  \\[1pt]
  Pure augmentation        & 81.38$\pm$4.91 & 77.56$\pm$3.57 & 86.24$\pm$2.95  \\[1pt]
  CFO comp. + channel aug.    & 81.59$\pm$2.48 & 81.98$\pm$1.52 & 91.80$\pm$2.11 \\
  \bottomrule
\end{tabular}
\end{small}
\end{subtable}
\vskip 1.8em
\begin{subtable}{\linewidth}
\caption{Performance when we use a single day for training, and then test on a different day.}
\centering
\begin{small}
\begin{tabular}{lcccccc}%
  \toprule
  \multirow{2}{*}{{Training Strategy}}
  & \multicolumn{3}{c}{Test time Augmentation} \\
  \cmidrule(l){2-4} 
  & None & 1 & 100 \\[1pt]
  \midrule
  No aug. or comp. & 5.47$\pm$4.49 & 2.72$\pm$1.07 & 3.90$\pm$2.75  \\[1pt]
  Pure augmentation        & 7.63$\pm$4.37  & 5.48$\pm$3.01 & 6.70$\pm$3.26  \\[1pt]
  CFO comp. + channel aug.    & 11.10$\pm$5.29 & 8.99$\pm$1.06 & 11.31$\pm$4.92 \\
  \bottomrule
\end{tabular}
\end{small}
\end{subtable}
\end{table*}

\begin{table*}
\small
\caption{Architecture details for CNNs used in ADS-B fingerprinting. Kernel sizes follow the notation [convolution size, input channels, output channels] for convolutional layers, and [input size, output size] for fully connected layers.}
\label{table:arch_adsb}
\vskip .5em
\begin{subtable}{\linewidth}
    \centering
    \caption{Complex-valued CNN}
    \label{table:arch_adsb_complex}
    \begin{tabularx}{.85\textwidth}{lXXXc}%
        \toprule
        {Layer} & {Kernel size} &  {Bias size} & {Output shape}
        & {No. of real parameters} \\
        \midrule
        Complex Input Layer & --  & -- & [320, 1] & -- \\[1pt]
        Complex Conv. &  [40, 1, 100] & -- & [15, 100] & 8000  \\[1pt]
        ModRelu &  -- & [100] & [15, 100] & 100  \\[1pt]
        Complex Conv.  &  [5, 100, 100] & -- & [11, 100] & 100000   \\[1pt]
        ModRelu & -- & [100] & [11, 100] & 100  \\[1pt]
        Absolute Value &  -- & -- & [11, 100] & -- \\[1pt]
        Global Average Pooling & --  & -- & [100] & -- \\[1pt]
        Real Fully Connected & [100, 100] & [100] & [100] & 10100 \\[1pt]
        Real Fully Connected  & [100, 100]  & [100] & [100] & 10100 \\
        \midrule
        Total & & & & 128400 \\        
        \bottomrule
    \end{tabularx}
\end{subtable}
\vskip 1.5em
\begin{subtable}{\linewidth}
    \centering
    \caption{Real (1x) CNN}
    \label{table:arch_adsb_real_1x}
   \begin{tabularx}{.85\textwidth}{lXXXc}%
        \toprule
        {Layer} & {Kernel size} &  {Bias size} & {Output shape}
        & {No. of real parameters} \\
        \midrule
        Stacked Re/Im Input Layer & -- & -- & [320, 2] & -- \\[1pt]
        Real Conv. & [40, 2, 100] & [100] & [15, 100] & 8100  \\[1pt]
        Real Conv.  &  [5, 100, 100] & [100] & [11, 100] & 50100   \\[1pt]
        Global Average Pooling & -- & -- & [100] & -- \\[1pt]
        Real Fully Connected & [100, 100] & [100] & [100] & 10100 \\[1pt]
        Real Fully Connected  & [100, 100] & [100] & [100] & 10100 \\
        \midrule
        Total & & & & 78400 \\          
        \bottomrule
    \end{tabularx}
\end{subtable}
\vskip 1.5em
\begin{subtable}{\linewidth}
    \centering
    \caption{Real (1.4x) CNN}
    \label{table:arch_adsb_real_14x}
    \begin{tabularx}{.85\textwidth}{lXXXc}%
        \toprule
        {Layer} & {Kernel size} &  {Bias size} & {Output shape}
        & {No. of real parameters} \\
        \midrule
        Stacked Re/Im Input Layer & -- & -- & [320, 2] & -- \\[1pt]
        Real Conv. & [40, 2, 140] & [140] & [15, 140] & 11340  \\[1pt]
        Real Conv.  & [5, 140, 140] & [140] & [11, 140] & 98140   \\[1pt]
        Global Average Pooling & -- & -- & [140] & -- \\[1pt]
        Real Fully Connected & [140, 100] & [100] & [100] & 14100 \\[1pt]
        Real Fully Connected  &  [100, 100] & [100] & [100] & 10100 \\
        \midrule
        Total & & & & 133680 \\  
        \bottomrule
    \end{tabularx}
\end{subtable}
\vskip 1.5em
\begin{subtable}{\linewidth}
    \centering
    \caption{Real (2x) CNN}
    \label{table:arch_adsb_real_2x}
    \begin{tabularx}{.85\textwidth}{lXXXc}%
        \toprule
        {Layer} & {Kernel size} &  {Bias size} & {Output shape}
        & {No. of real parameters} \\
        \midrule
        Stacked Re/Im Input Layer & -- & -- & [320, 2] & -- \\[1pt]
        Real Conv. & [40, 2, 200] &  [200] & [15, 200] & 16200  \\[1pt]
        Real Conv.  & [5, 200, 200] & [200] &  [11, 200] & 200200   \\[1pt]
        Global Average Pooling & -- &  -- & [200] & -- \\[1pt]
        Real Fully Connected & [200, 100] & [100]  & [100] & 20100 \\[1pt]
        Real Fully Connected  & [100, 100] &  [100] &  [100] & 10100 \\
        \midrule
        Total & & & & 246600 \\  
        \bottomrule
    \end{tabularx}
\end{subtable}
\end{table*}

\begin{table*}
\centering
\small
\caption{Architecture details for CNNs used in WiFi fingerprinting. Kernel sizes follow the notation [convolution size, input channels, output channels] for convolutional layers, and [input size, output size] for fully connected layers.}
\label{table:arch_wifi}
\vskip .5em
\begin{subtable}{\linewidth}
    \centering
    \caption{Complex-valued CNN}
    \label{table:arch_wifi_complex}
    \begin{tabularx}{.85\textwidth}{lXXXc}%
        \toprule
        {Layer} & {Kernel size} &  {Bias size}& {Output shape}
        & {No. of real parameters} \\
        \midrule
        Complex Input Layer & -- & -- & [3200, 1] & -- \\[1pt]
        Complex Conv. & [200, 1, 100] & [100] &  [31, 100] & 40200  \\[1pt]
        ModRelu & -- & [100] & [31, 100] & 100  \\[1pt]
        Complex Conv.  & [10, 100, 100]  & -- & [22, 100] & 200200   \\[1pt]
        ModRelu & -- & [100] & [22, 100] & 100  \\[1pt]
        Absolute Value & -- & -- & [22, 100] & -- \\[1pt]
        Real Fully Connected & [100, 100] & [100] & [22, 100] & 10100 \\[1pt]
        Real Fully Connected  & [100, 100] &  [100] & [22, 100] & 10100 \\[1pt]
        Global Average Pooling & -- & -- & [100] & -- \\[1pt]
        Real Fully Connected & [100, 19]  & [19] & [19] & 1919 \\
        \midrule
        Total & & & & 262719 \\   
        \bottomrule
    \end{tabularx}
\end{subtable}
\vskip 1.5em
\begin{subtable}{\linewidth}
    \centering
    \caption{Real (1x) CNN}
    \label{table:arch_wifi_real_1x}
    \begin{tabularx}{.85\textwidth}{lXXXc}%
        \toprule
        {Layer} & {Kernel size}&  {Bias size} & {Output shape}
        & {No. of real parameters} \\
        \midrule
        Stacked Re/Im Input Layer & -- & -- &  [3200, 2] & -- \\[1pt]
        Real Conv. & [200, 2, 100] & [100] &  [31, 100] & 40100  \\[1pt]
        Real Conv.  & [10, 100, 100]  & [100] & [22, 100] & 100100   \\[1pt]
        Real Fully Connected & [100, 100]  & [100] & [22, 100] & 10100 \\[1pt]
        Real Fully Connected  &  [100, 100] & [100] & [22, 100] & 10100 \\[1pt]
        Global Average Pooling & -- & -- & [100] & -- \\[1pt]
        Real Fully Connected & [100, 19] & [19] & [19] & 1919 \\
        \midrule
        Total & & & & 162319 \\   
        \bottomrule
    \end{tabularx}
\end{subtable}
\vskip 1.5em
\begin{subtable}{\linewidth}
    \centering
    \caption{Real (1.4x) CNN}
    \label{table:arch_wifi_real_14x}
    \begin{tabularx}{.85\textwidth}{lXXXc}%
        \toprule
        {Layer} & {Kernel size}&  {Bias size} & {Output shape}
        & {No. of real parameters} \\
        \midrule
        Stacked Re/Im Input Layer & -- & -- & [3200, 2] & -- \\[1pt]
        Real Conv. & [200, 2, 140] & [140] & [31, 140] & 56140  \\[1pt]
        Real Conv.  & [10, 140, 140] & [140] & [22, 140] & 196140   \\[1pt]
        Real Fully Connected & [140, 100] & [100] & [22, 100] & 14100 \\[1pt]
        Real Fully Connected  &  [100, 100] & [100] & [22, 100] & 10100 \\[1pt]
        Global Average Pooling & -- & -- & [100] & -- \\[1pt]
        Real Fully Connected & [100, 19] & [19] & [19] & 1919 \\
        \midrule
        Total & & & & 278399 \\
        \bottomrule
    \end{tabularx}
\end{subtable}
\vskip 1.5em
\begin{subtable}{\linewidth}
    \centering
    \caption{Real (2x) CNN}
    \label{table:arch_wifi_real_2x}
    \begin{tabularx}{.85\textwidth}{lXXXc}%
        \toprule
        {Layer} & {Kernel size}&  {Bias size} & {Output shape}
        & {No. of real parameters} \\
        \midrule
        Stacked Re/Im Input Layer & -- & -- & [3200, 2] & -- \\[1pt]
        Real Conv. & [200, 2, 200] & [200] & [31, 200] & 80200  \\[1pt]
        Real Conv.  & [10, 200, 200] & [200] &  [22, 200] & 400200   \\[1pt]
        Real Fully Connected & [200, 100] & [100] & [22, 100] & 20100 \\[1pt]
        Real Fully Connected  &  [100, 100] & [100] & [22, 100] & 10100 \\[1pt]
        Global Average Pooling & -- & -- & [100] & -- \\[1pt]
        Real Fully Connected & [100, 19] & [19] & [19] & 1919 \\
        \midrule
        Total & & & & 512519 \\
        \bottomrule
    \end{tabularx}
\end{subtable}
\end{table*}

\end{appendices}
\end{document}